\documentclass[11pt,preprint]{emulateapj}
\usepackage{lscape}
\def\dlambda{$\lambda\lambda$}
\def\kms{km~s$^{-1}$} 

\slugcomment{Submitted to the Astrophysical J.\ 2009 April 27; accepted 2009 December 15}

\shorttitle{Late-Time [O I] Emission Peaks}
\shortauthors{Milisavljevic et al.}

\begin{document}

\title{Doublets and Double Peaks: Late-Time [O I] \dlambda 6300, 6364 
Line Profiles \\ of Stripped-Envelope, Core-Collapse Supernovae}

\author{Dan Milisavljevic\altaffilmark{1},
        Robert A.~Fesen\altaffilmark{1},
	Christopher L.~Gerardy\altaffilmark{2},
        Robert P.~Kirshner\altaffilmark{3},
        Peter Challis\altaffilmark{3} }

\altaffiltext{1}{6127 Wilder Lab, Department of Physics \& Astronomy, Dartmouth
                 College, Hanover, NH 03755, USA}
\altaffiltext{2}{Department of Physics, Florida State University, Tallahassee, FL 
32306, USA}
\altaffiltext{3}{Harvard-Smithsonian Center for Astrophysics, 60 Garden Street,
                 Cambridge, MA 02138, USA} 

\begin{abstract}
We present optical spectra of SN~2007gr, SN~2007rz, SN~2007uy, SN~2008ax, and
SN~2008bo obtained in the nebular phase when line profiles can lead to
information about the velocity distribution of the exploded cores. We compare
these to 13 other published spectra of stripped-envelope core-collapse
supernovae (Type IIb, Ib, and Ic) to investigate properties of their
double-peaked [\ion{O}{1}] \dlambda 6300, 6364 emission.  These 18 supernovae
are divided into two empirical line profile types: (1) profiles showing two
conspicuous emission peaks nearly symmetrically centered on either side of 6300
\AA\ and spaced $\approx$64~\AA\ apart, close to the wavelength
separation between the [\ion{O}{1}] \dlambda 6300, 6364 doublet lines, and
(2) profiles showing asymmetric [\ion{O}{1}] line profiles consisting of a
pronounced emission peak near 6300~\AA \ plus one or more blueshifted emission
peaks.  Examination of these emission profiles, as well as comparison with
profiles in the lines of [\ion{O}{1}] $\lambda$5577, \ion{O}{1} $\lambda$7774,
and \ion{Mg}{1}] $\lambda$4571, leads us to conclude that neither type of [\ion{O}{1}]
double-peaked profile is necessarily the signature of emission from front
and rear faces of ejecta arranged in a toroidal disk or elongated shell
geometry as previously suggested. We propose possible alternative
interpretations of double-peaked emission for each profile type, test their
feasibility with simple line-fitting models, and discuss their strengths and
weaknesses.  The underlying cause of the observed predominance of blueshifted
emission peaks is unclear, but may be due to internal scattering or dust
obscuration of emission from far side ejecta.  
\end{abstract}


\section{Introduction}

Analyses of emission-line profiles of ejecta-tracing elements in
stripped-envelope, core-collapse supernovae (CCSNe) during the ``nebular
phase'' several months after outburst probe the chemical and kinematic
properties of the metal-rich ejecta, thereby yielding clues about CCSN explosion
dynamics and geometry.  Particular attention has been paid to studying the line
profiles of oxygen, magnesium, and calcium as these elements are among the
strongest lines $100-200$ days post-outburst \citep{Fransson89,Filippenko97}.
Stripped-envelope SNe lacking strong H$\alpha$ emission (i.e., Types IIb, Ib,
and Ic) can be particularly informative because details of the interior regions
are not obscured by the hydrogen envelope surrounding progenitors of Type II
SNe. 
 
Late-time spectra of stripped CCSNe have shown double-peaked line profiles
in [\ion{O}{1}] \dlambda 6300, 6364 emission to be a relatively common
phenomenon, suggesting a possible unifying characteristic across types IIb, 
Ib, and Ic.  \citet{Maeda08} found double-peaked [\ion{O}{1}] \dlambda 6300,
6364 profiles in 40\% of a sample of 18 stripped CCSNe, while \citet{Modjaz08a}
reported finding double-peaked [\ion{O}{1}] profiles in three of eight objects
studied. Such spectra show two conspicuous emission peaks, often symmetrically
situated around a trough centered near 6300 \AA\ and, when present, persist
over the full observing period.  

\begin{figure*}[htp!] \centering
\includegraphics[width=0.50\linewidth]{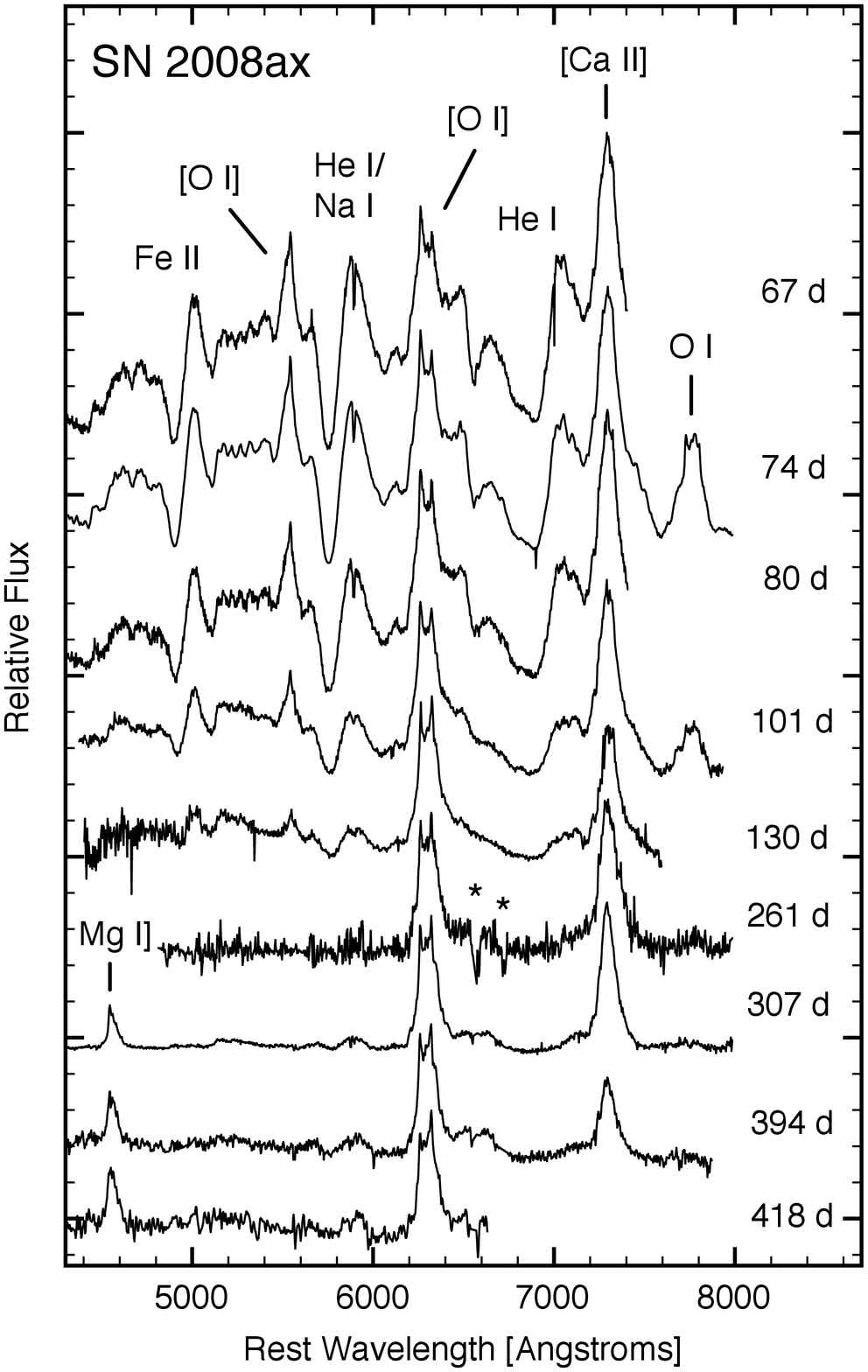}
\includegraphics[width=0.469\linewidth]{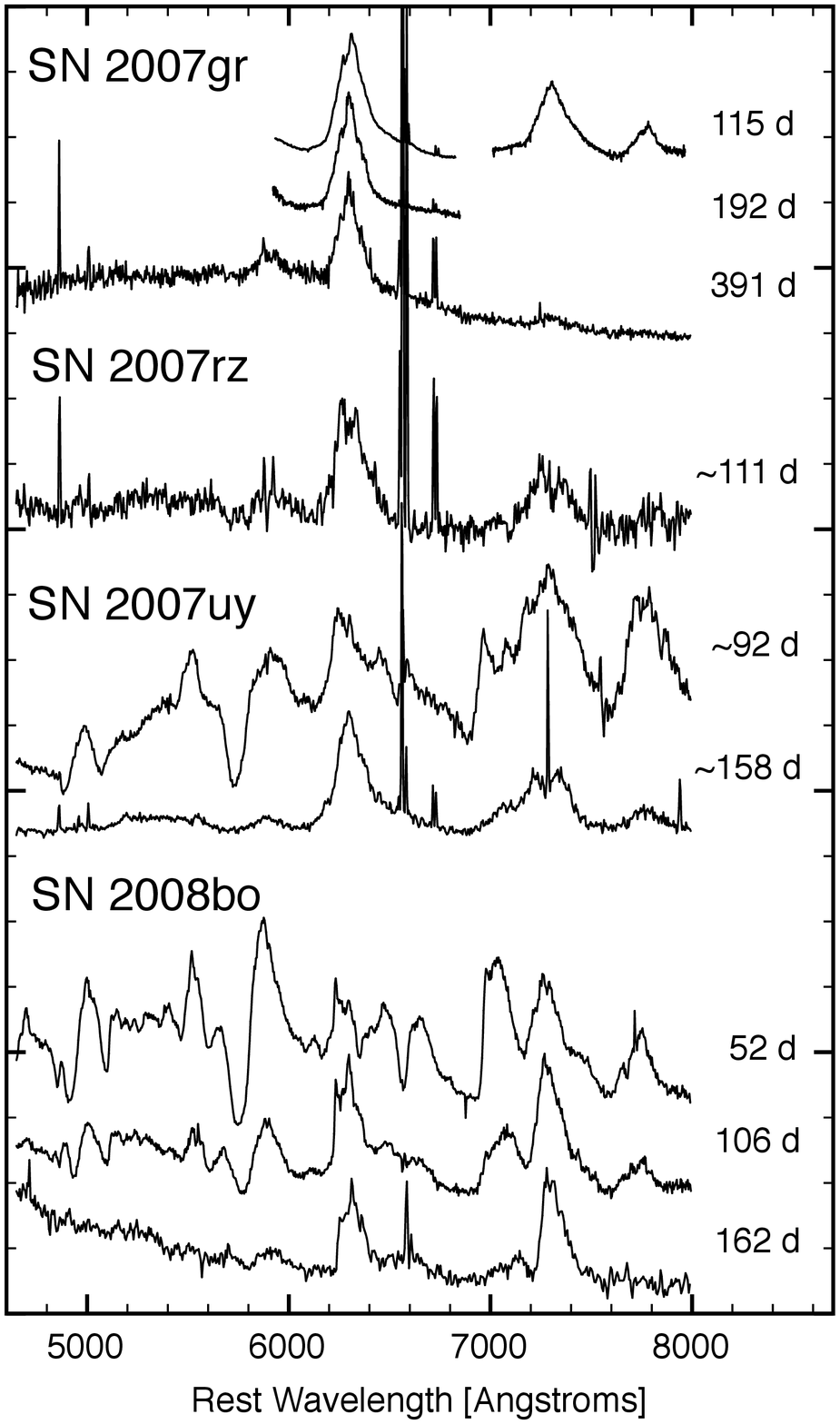} 
\caption{{\it Left panel}:
optical spectra of SN~2008ax spanning its evolution into the nebular
phase of emission.  Asterisks (*) mark artificial features introduced during
the reduction process from oversubstracted nebular lines emitted by 
coincident \ion{H}{2} regions.  {\it Right panel}: optical spectra of
SN~2007gr, SN~2007rz, SN~2007uy, and SN~2008bo.  See Table 1 for details of all
observations.} 
\end{figure*}

Double-peaked emission-line profiles deviate from the single-peaked profile
expected from a spherically symmetric source and this has been interpreted in
many cases as evidence for aspherically distributed debris having a torus or
disk-like geometry. \citet{Maeda02} first predicted that double-peaked
[\ion{O}{1}] line profiles could be associated with a torus of O-rich ejecta.
\citet{Mazzali05} observed a double-peaked [\ion{O}{1}] profile in the Type Ic
broad-lined SN~2003jd, and interpreted it as emission originating from a torus
of O-rich debris perpendicular to a high-velocity jet in a gamma-ray burst
(GRB) model.  \citet{Maeda08} reported observing double-peaked profiles in the
optical spectra of approximately 7 out of 18 CCSNe and was able to model the
profiles using an aspherical, torus-like geometry of O-rich ejecta. They
concluded that the observed [\ion{O}{1}] emission could be either single- or
double-peaked depending on the viewing angle being either perpendicular to or
along the torus plane.  \citet{Modjaz08a} also interpreted double-peaked
[\ion{O}{1}] profiles in several CCSN spectra and found them consistent with a
torus-like distribution of SN debris generated by the explosion physics, but
not necessarily a GRB-like jet/torus structure.  

Here we present observations and analyses to help characterize the [\ion{O}{1}]
\dlambda 6300, 6364 emission profile in late-time optical spectra of
stripped-envelope CCSNe.  In Sections $2$ and $3$ we present new low- to
moderate-resolution optical spectra of five recent stripped-envelope CCSNe
obtained $2-14$ months after optical maximum. We then compare the [\ion{O}{1}]
emission-line profiles of these and 13 other published late-time spectra in
Sections $4$ and $5$.  In Section $6$ we summarize our findings, concluding in
part that double-peaked [\ion{O}{1}] line profiles are not necessarily the
signature of emission from front and rear faces of O-rich ejecta arranged in a
toroidal geometry as previously suggested. 

\section{Observations}

Low-dispersion optical spectra were obtained with the 6.5~m MMT and the 1.5~m
FLWO telescopes at Mt.\ Hopkins in Arizona, and the 2.4~m Hiltner telescope at
the MDM Observatory on Kitt Peak, Arizona. MMT observations used the Blue
Channel spectrograph \citep{Schmidt89} employing a 1$''$ wide slit and a 300
lines mm$^{-1}$ 4800 \AA\ blaze grating.  Spectra typically spanned $3500-8000$
\AA\ with a resolution of $\approx 7$ \AA.  FLWO observations employed the FAST
spectrograph \citep{Fabricant98} with a 300 lines mm$^{-1}$ grating and
a standard slit width of 3$''$, yielding spectra covering $4000-7400$ \AA\ with
7~\AA\ resolution.

\begin{deluxetable*}{llclcccl}
\centering
\tablecaption{Summary of observations}
\tablecolumns{8}
\tabletypesize{\footnotesize}
\tablewidth{0pt}
\tablehead{\colhead{Supernova/}                &
	   \colhead{Host Galaxy}               &
	   \colhead{Redshift\tablenotemark{c}} &  
           \colhead{Observation}               &
           \colhead{Epoch}                     &
           \colhead{Telescope/}                &
	   \colhead{Spec.\ Res.}               &
	   \colhead{Exp.\ Time}                \\
           \colhead{Type}                      &
	   \colhead{}                          &
           \colhead{\kms}	               &
           \colhead{Date}                      &
           \colhead{(days)}                    &
           \colhead{Instrument}                &
	   \colhead{(\AA)}                     &
           \colhead{(s)}}                     
\startdata
SN 2007gr (Ic)  & NGC 1058 & 503 & 2007 Dec 21 & 115\tablenotemark{\ } & MDM/CCDS         & 
2 & $3000 \times 3$\\
                & & & 2008 Mar 07 & 192\tablenotemark{\ } & MDM/CCDS         & 
2 & $1800 \times 2$\\
                & & & 2008 Sep 22 & 391                 & MDM/Modspec      & 
6 & $3000 \times 3$\\
SN 2007rz (Ic)  & NGC 1590 & 4023 & 2008 Apr 01 & 111\tablenotemark{a}& MMT/Blue Channel & 
7\tablenotemark{b} & $600 \times 3$ \\
SN 2007uy (Ib)  & NGC 2770 & 1874 & 2008 Apr 01 & 92\tablenotemark{a} & MMT/Blue Channel &
7\tablenotemark{b} & $600 \times 4$ \\ 
	& & & 2008 June 06 & 158\tablenotemark{a}&
MMT/Blue Channel & 7\tablenotemark{b} & $900$ \\
SN 2008ax (IIb) & NGC 4490 & 565\tablenotemark{d} & 2008 May 30 &
67                  & FLWO/FAST         & 10 & $1800$        \\
& & & 2008 Jun 06 & 74& MMT/Blue Channel & 7  & $900$         \\ 
& & & 2008 Jun 12 & 80& FLWO/FAST        & 11 & $1200$        \\ 
& & & 2008 Jul 03 &101& MDM/CCDS         & 11 & $500$         \\
& & & 2008 Aug 01 &130& MDM/Modspec      & 6  & $600$         \\
& & & 2008 Dec 10 &261& MDM/CCDS         & 11 & $2700 \times 2$ \\ 
& & & 2009 Jan 25 &307& MMT/Blue Channel & 7  & $1800$ \\
& & & 2009 Apr 22\tablenotemark{e} &394& MDM/CCDS          & 11 & $2700 \times 4$\\
& & & 2009 May 16 &418& MDM/CCDS          & 11 & $2700$ \\
SN 2008bo (IIb)  & NGC 6643 & 1484 & 2008 Jun 06 & 52 & MMT/Blue Channel & 7  & $900$  \\ 
& & & 2008 Jul 30 &106& MMT/Blue Channel & 7  & $900$         \\ 
& & & 2008 Sep 24 &162& MDM/Mark III     & 12 & $2000 \times 2$
\enddata
\tablenotetext{a}{Epoch with respect to discovery date.}
\tablenotetext{b}{Presented spectrum smoothed with a 3 pixel boxcar function.}
\tablenotetext{c}{Recessional velocity inferred from narrow H$\alpha$ line unless otherwise noted.}
\tablenotetext{d}{No adjacent H$\alpha$ detected; NED recessional velocity of host galaxy used.}
\tablenotetext{e}{Presented spectrum is average of two nights of observations, 2009 April 22 and 23.}
\end{deluxetable*} 

MDM observations used three different spectrographs. A Boller \& Chivens CCD
spectrograph (CCDS) was used with a north-south 1.2$\arcsec \times$ 5$\arcmin$
slit and either a 150 lines mm$^{-1}$ 4700 \AA\ blaze grating yielding $\approx
10$ \AA\ resolution, or a 600 lines mm$^{-1}$ 4700 \AA\ blaze yielding 2 \AA\
resolution. The Modular Spectrograph was used in combination with the SITe 2K
Echelle CCD detector with a 1.2$\arcsec \times$ 5$\arcmin$ slit and a 600 lines
mm$^{-1}$ 5000 \AA\ blaze grating.  Resulting spectra spanned $4500-7500$ \AA\
with a resolution of 6 \AA. The Mark III spectrograph in combination with a SITe
1K CCD detector (`Templeton') was used with a 300 lines mm$^{-1}$ 6400 \AA\
blaze grism yielding spectra of 12 \AA\ resolution.  Details of all
observations including dates and exposure times are provided in Table 1.

All spectra were reduced and calibrated employing standard techniques in
IRAF\footnote{IRAF is distributed by the National Optical Astronomy
Observatories, which are operated by the Association of Universities for
Research in Astronomy, Inc., under cooperative agreement with the National
Science Foundation.} and our own IDL routines (see, e.g.,
\citealt{Matheson08}). Narrow host galaxy emission features, telluric
absorptions, and obvious cosmetic defects have been removed from all spectra.
All reported wavelengths are in the rest frame of the supernovae as determined
by the recessional velocity inferred from narrow H$\alpha$ lines observed in
local \ion{H}{2} regions.  When no such lines were available, host galaxy
heliocentric recessional velocities were retrieved on-line from the NASA/IPAC
Extragalactic Database (NED)\footnote{http://nedwww.ipac.caltech.edu/.}.
Reported epochs are with respect to published dates of maximum optical
brightness noted below, otherwise they are estimated with respect to discovery
dates. 

\section{Results}

In Figure 1 we present low- to moderate-resolution late-time optical
spectra of five stripped-envelope CCSNe observed 2007--2009 at epochs spanning
evolution into the nebular phase when emission is dominated by forbidden
transitions.  The left panel presents spectra of SN 2008ax, and the right panel
spectra of SN~2007gr, SN~2007rz, SN~2007uy, and SN~2008bo. The following is a
brief description of these data with particular emphasis on the emission-lines
of ejecta-tracing elements oxygen, magnesium, and calcium which will be
discussed in greater depth in Section $4$.

\subsection{SN 2008ax}

SN 2008ax in NGC~4490 was discovered by \citet{Mostardi08} on 2008 March 3 and
classified spectroscopically as a Type IIb SN \citep{Chornock08}.
Extensive spectra and photometric monitoring of the SN and investigation
of pre-explosion {\sl Hubble Space Telescope} ({\sl HST}) images show it
consistent with the explosion of a young Wolf-Rayet progenitor star of WNL type
\citep{Crockett08b,Pastorello08}.  The SN reached maximum brightness
$m_{V} = 13.5$ on 2008 March 24 \citep{Pastorello08}.

Our late-time spectra (Figure 1, {\it left panel}) show early P Cygni features
attributable to  \ion{Fe}{2} lines around 5000 \AA, blended \ion{He}{1}
$\lambda$5876 and \ion{Na}{1} $ \lambda\lambda$5890, 5896, and \ion{He}{1}
$\lambda$7065 that decline in strength with  time. Emission from forbidden
lines [\ion{O}{1}] $\lambda\lambda$6300, 6364 and [\ion{Ca}{2}]
$\lambda\lambda$7291,7324 remains strong at all epochs. We interpret the
relatively strong emission peaked at 4544 \AA\ observable on days 307, 394, and
418 with \ion{Mg}{1}] $\lambda$4571.  Oxygen emission from [\ion{O}{1}]
$\lambda$5577 and \ion{O}{1} $\lambda$7774 decline in strength relative to
[\ion{O}{1}] \dlambda 6300, 6364. Two conspicuous narrow emission peaks around
6260 \AA\ and 6323 \AA\ ($-1800$ \kms\ and $+1200$ \kms, with respect to
6300~\AA) are noticeable in the [\ion{O}{1}] profile from the earliest spectrum
on day 67 and persist until our last observation on day 418.

In Figure 2, we show these same spectra enlarged around emission-lines of
interest to better see temporal changes in the observed features. The
[\ion{O}{1}] $\lambda$5577 line (Figure 2, {\it left}) shows only a blue peak,
with nearly all of the emission profile falling blueward of zero velocity.  The
weighted center of this emission shifts to smaller velocities with time, until
day 261 when no significant [\ion{O}{1}] $\lambda$5577 emission is detected.
On the other hand, the [\ion{O}{1}] $\lambda\lambda$6300, 6364 lines ({\it
middle}) show two peaks roughly symmetric about 6300 \AA.  The relative
strengths of the blue and red peaks change noticeably, with the blue peak
stronger at early times but gradually becoming weaker at later epochs. The
[\ion{Ca}{2}] $\lambda\lambda$7291,7324 profile ({\it right}) is only modestly
blueshifted and shows no change in the weighted line center over time.  

\begin{figure*}
\centering
\includegraphics[width=0.3\linewidth]{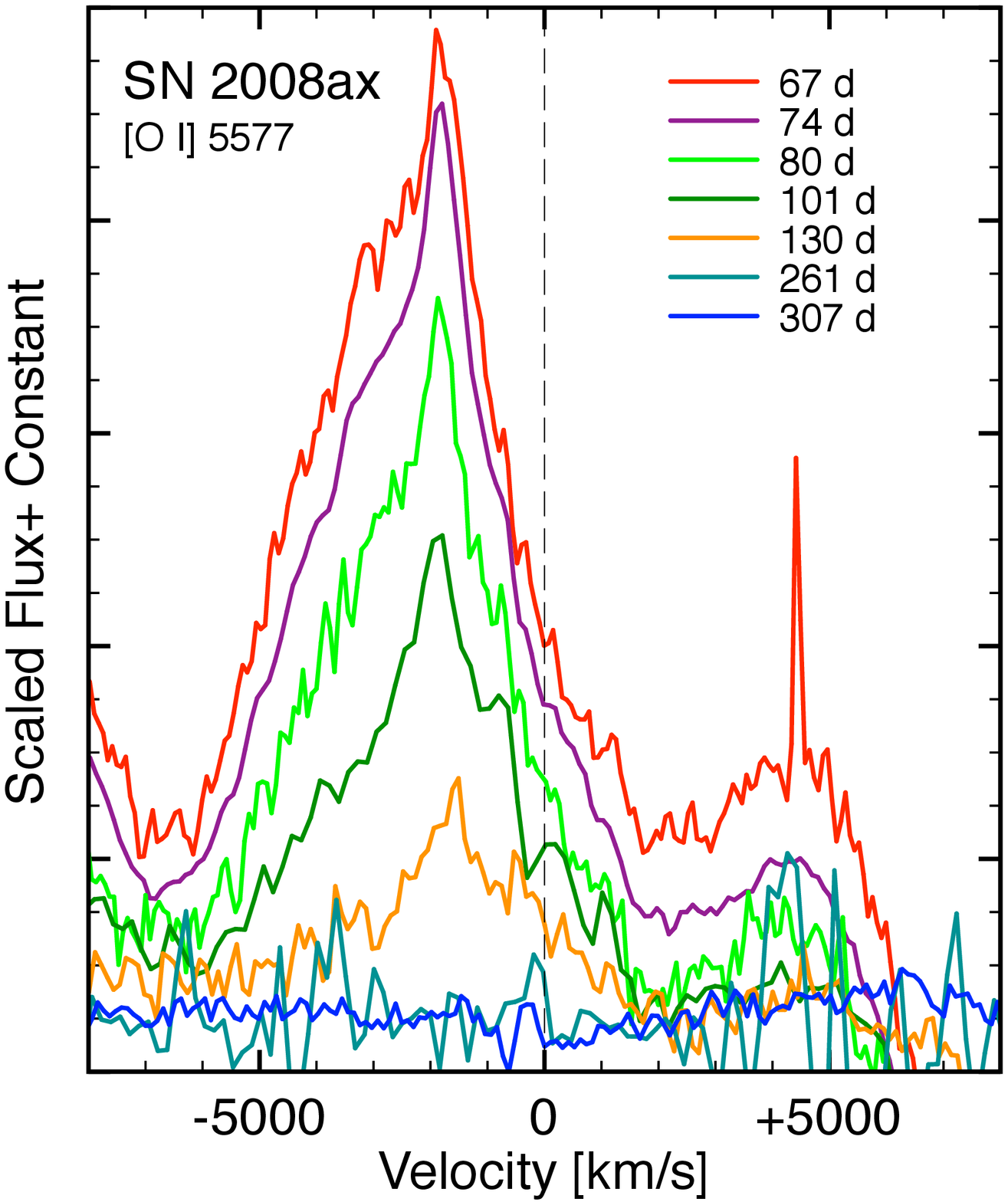}
\includegraphics[width=0.3\linewidth]{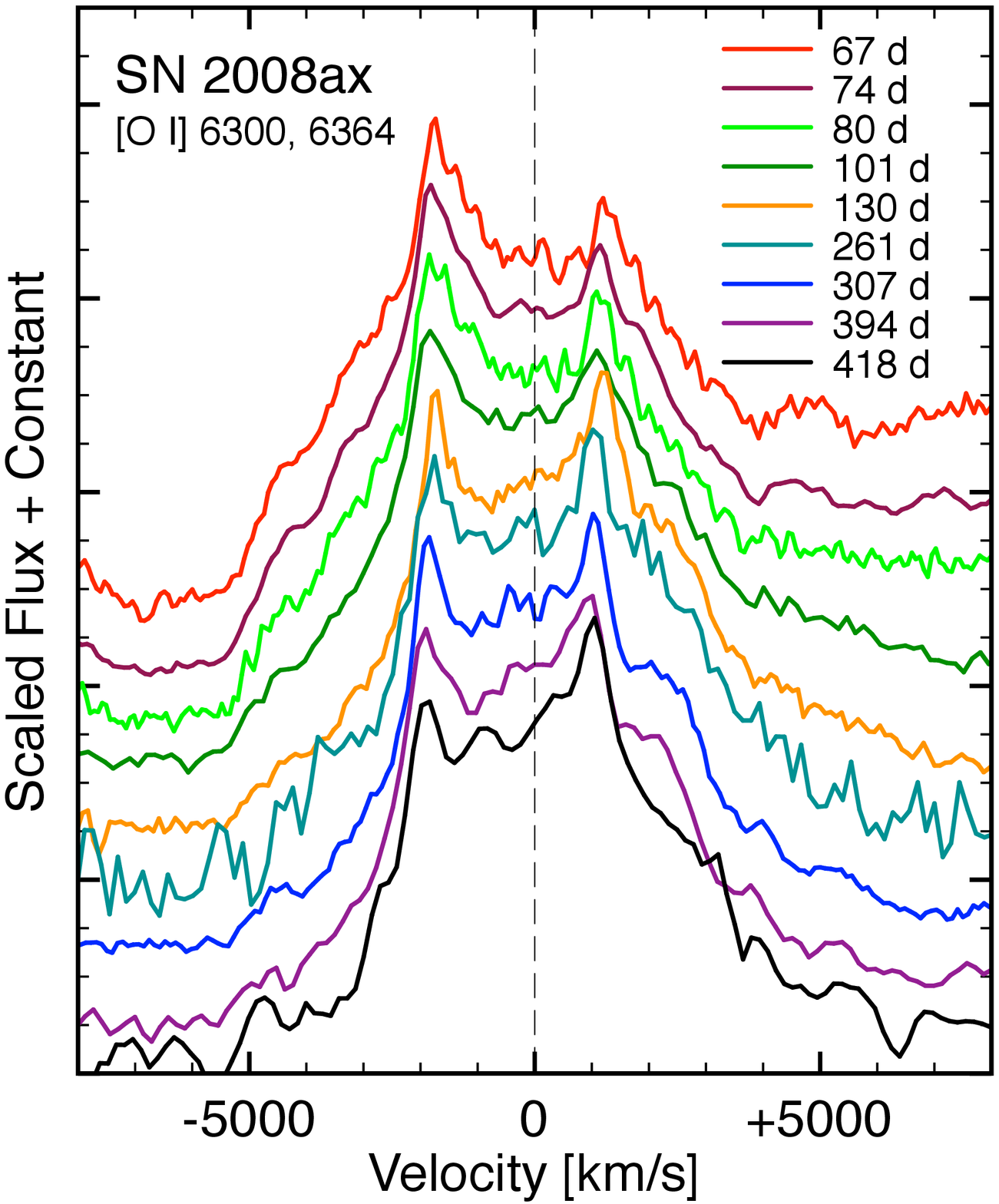}
\includegraphics[width=0.3\linewidth]{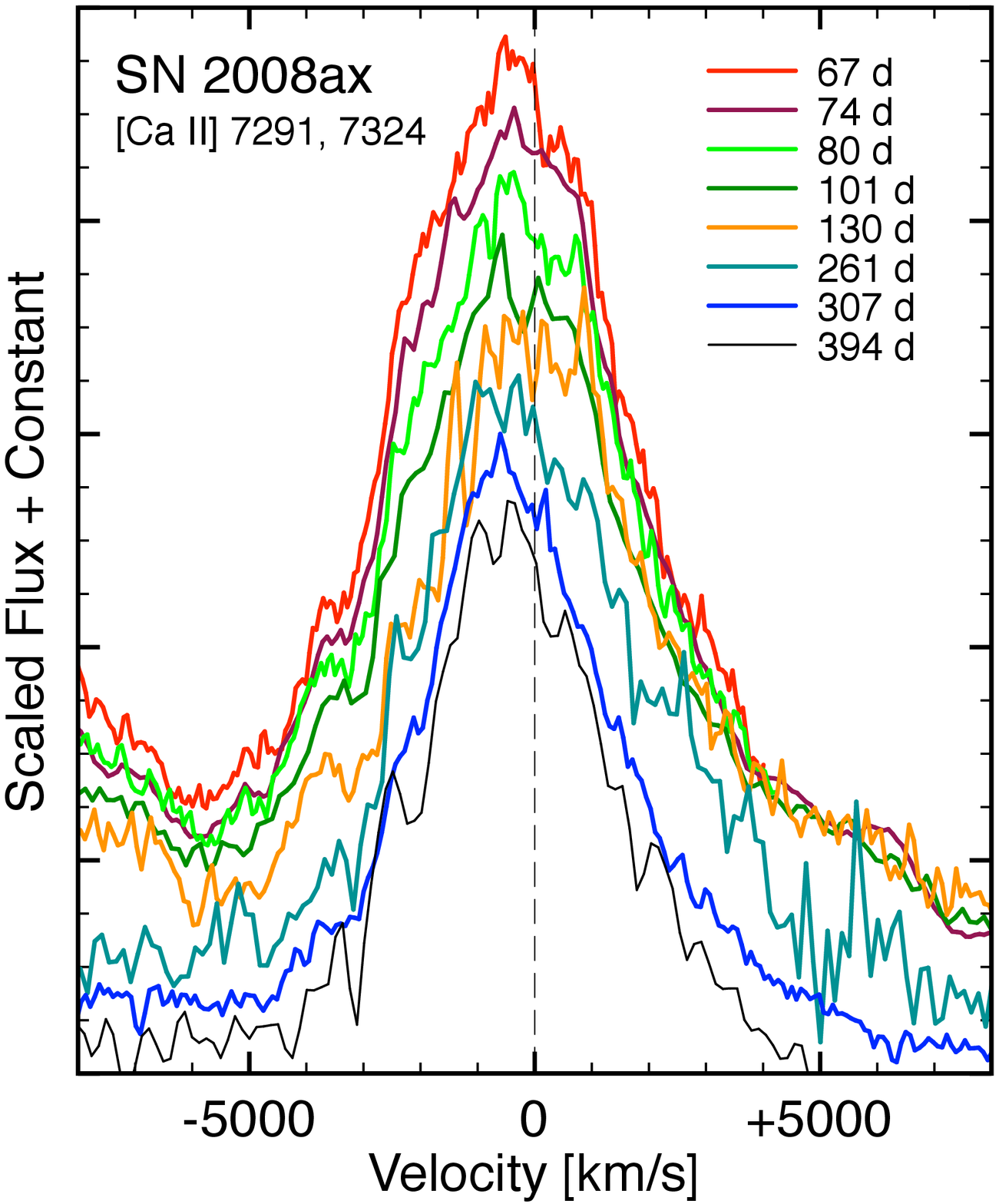}
\caption{Velocity line profiles of [\ion{O}{1}] $\lambda$5577, [\ion{O}{1}]
\dlambda 6300, 6364, and [\ion{Ca}{2}] \dlambda 7291, 7324.  The dashed line
marks zero velocity with respect to 5577, 6300, and 7306 \AA, respectively.
Time progresses from top to bottom.} 
\end{figure*}

\begin{figure}[ht!p] 
\centering 
\includegraphics[width=0.95\linewidth]{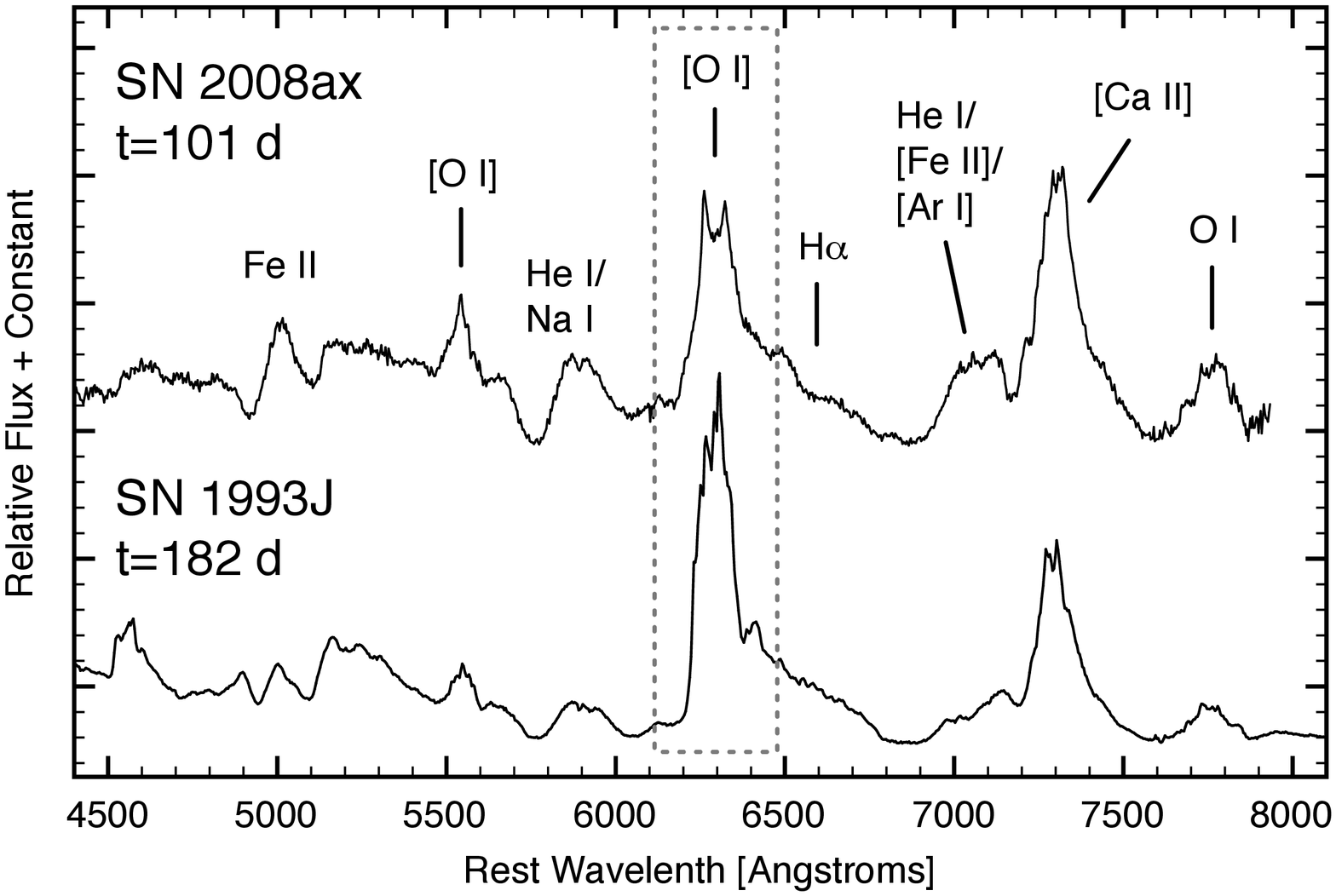} \\ 
\includegraphics[width=0.95\linewidth]{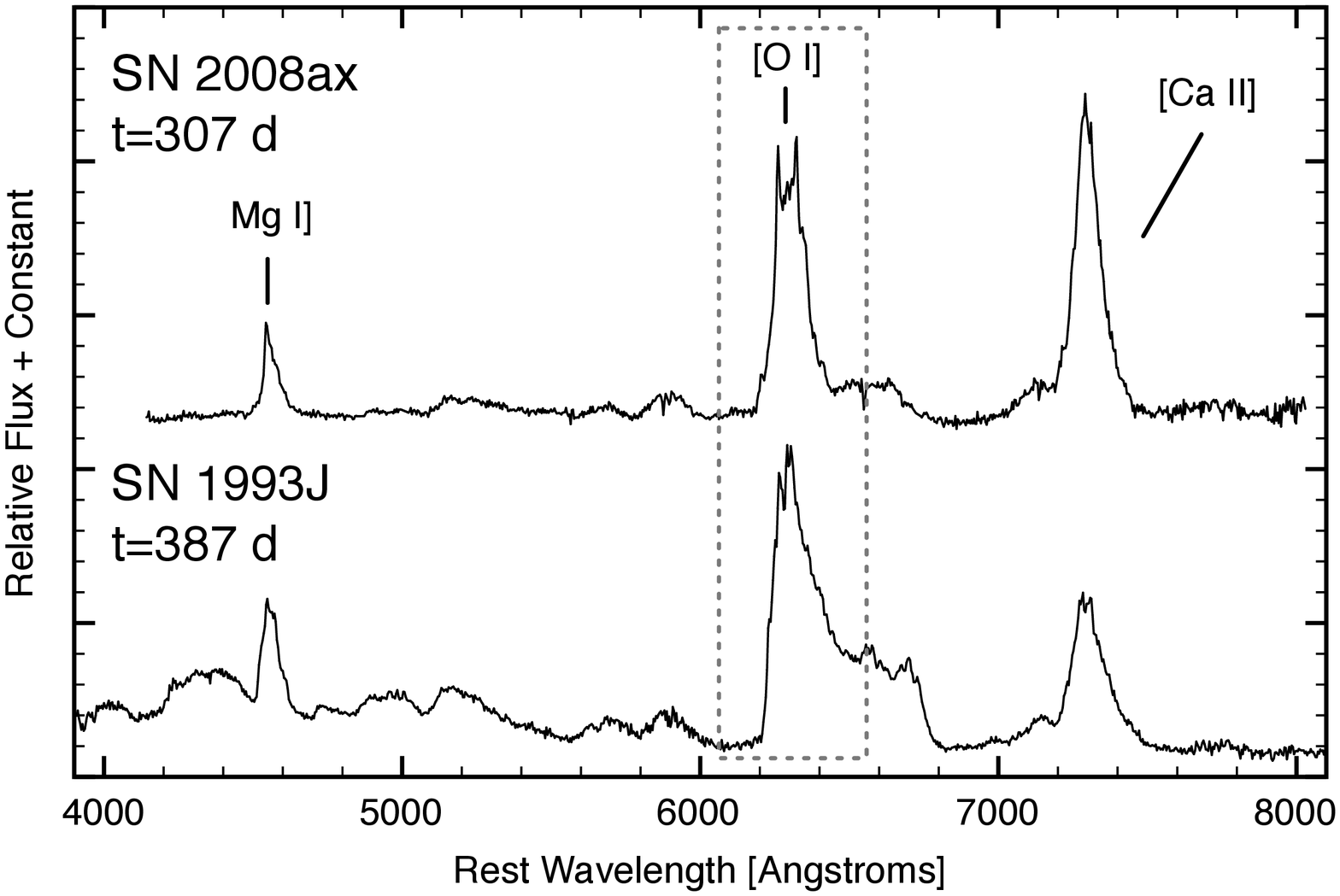} 
\caption{Late-time optical spectra of SN~2008ax compared to SN~1993J.  
Most spectral features are similar except for emission around [\ion{O}{1}]
\dlambda 6300, 6364.  The dashed box highlights SN~2008ax's double-peaked
[\ion{O}{1}] emission and SN~1993J's single-peaked emission.  Data for
SN~1993J are from \citet{Matheson00a,Matheson00b}.} 
\end{figure}

Finally, we note that although SN 2008ax shares many spectral features with the
SN IIb event SN~1993J, SN~2008ax displays a clear and prominent double-peaked
[\ion{O}{1}] \dlambda 6300, 6364 emission profile which SN~1993J did not.  In
Figure 3, we show SN 2008ax's day 101 and day 307 spectra alongside optical
spectra of SN~1993J taken on days 182 and 387, measured from date of maximum
brightness.  Data for SN~1993J are from
\citet{Matheson00a,Matheson00b}.\footnote{Retrieved on-line at
http://www.noao.edu/ noao/ \\ staff/matheson/spectra.html.}

\subsection{SN 2007gr}

SN 2007gr in NGC~1058 was discovered on 2007 August 15 by \citet{Madison07} as
part of the Lick Observatory Supernova Search. It was spectroscopically
identified as a Type Ib/c by \citet{Chornock07} and later confirmed as a Type
Ic by \citet{Valenti08a}.   It reached maximum brightness $m_{R} = 12.8$ on
2007 August 28 \citep{Valenti08a}.  Pre-explosion {\sl HST} WFPC2 and
ground-based {\sl K}-band images marginally favor a progenitor star from a $7.0
\pm 0.5$  Myr cluster having a turn-off mass of $28 \pm 4$ M$_{\sun}$
\citep{Crockett08a}. Extensive optical and near-infrared observations
from days 5 to 415 are presented by \citet{Hunter09}.

Our moderate-resolution spectrum of SN~2007gr on day 115 (Figure 1, {\it right
panel}) shows broad (HWZI $\ga 6000$ km s$^{-1}$) [\ion{O}{1}] \dlambda 6300,
6364 emission strongly peaked around 6300 \AA, along with [\ion{Ca}{2}]
\dlambda 7291, 7324, and \ion{O}{1} $\lambda$7774 emission peaked at 7291 and
7770 \AA, respectively.  A minor emission peak around 6260 \AA\ ($-1900$ km
s$^{-1}$) is observed in the [\ion{O}{1}] \dlambda 6300, 6364 profile and the
high signal-to-noise ratio (S/N) of the spectra ensures that this feature is
real.  By day 391 [\ion{O}{1}] 6300, 6364 emission remains relatively strong
and chiefly single-peaked, and the minor blueshifted peak is no longer
detected.  

\subsection{SN 2007rz}

SN 2007rz in NGC~1590 was discovered on 2007 December 12 by \citet{Parisky07}
as part of the Lick Observatory Supernova Search and classified as a Type Ic by
\citet{Morrell07}.  Our $\sim 111$~d spectrum of SN~2007rz (Figure 1, {\it right
panel}) shows weak [\ion{O}{1}] emission roughly symmetric about 6300 \AA\ with
two peaks at 6264 and 6330 \AA\ (approximately $-1700$ and $1400$ \kms,
respectively).  Broad [\ion{Ca}{2}] \dlambda 7291, 7324 emission centered
around 7305 \AA\ is also observed.

\begin{figure} 
\centering 
\includegraphics[width=0.8\linewidth]{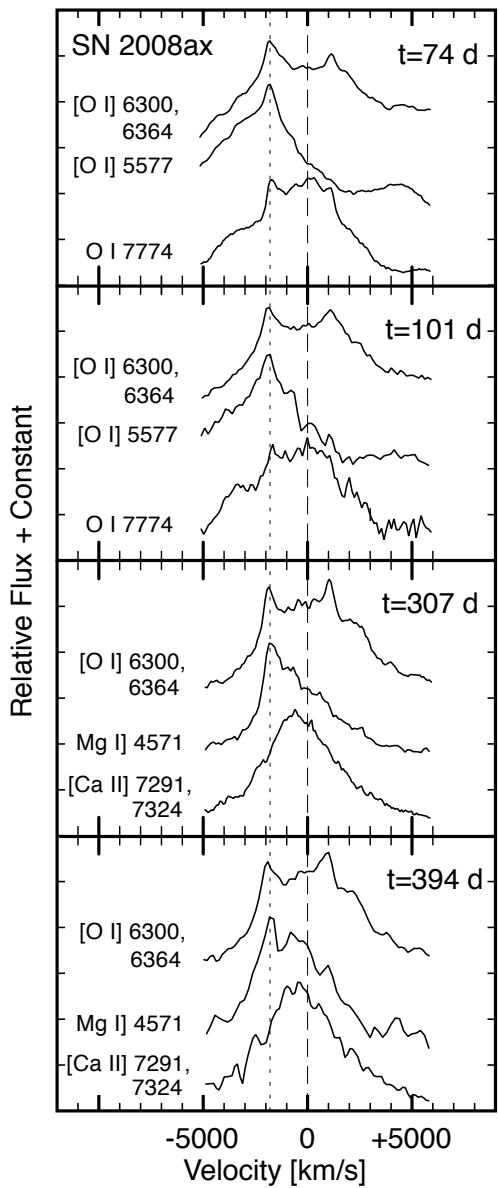}
\caption{Comparing emission-line profiles of [\ion{O}{1}] \dlambda 6300, 6364,
[\ion{O}{1}] $\lambda$5577, \ion{O}{1} $\lambda$7774, [\ion{Ca}{2}] \dlambda
7291, 7324, and \ion{Mg}{1}] $\lambda$4571 in SN~2008ax at four epochs.  The
heavy dashed line marks zero velocity with respect to wavelengths 6300, 5577,
7774, 7306, and 4571 \AA, respectively.  The fainter dashed line highlights
blueshifted peaks around $-1800$ \kms\ shared across [\ion{O}{1}] \dlambda
6300, 6364, [\ion{O}{1}] $\lambda$5577, and \ion{Mg}{1}] $\lambda$4571.}
\end{figure}

\subsection{SN 2007uy}

SN 2007uy in NGC~2770 was discovered 2007 December 31 by Y.\ Hirose
\citep{Nakano08} and classified as a Type Ib by \citet{Blondin08}.  The
SN has not quite reached a fully nebular phase in our day $\sim 92$
spectrum (Figure 1, {\it right panel}) and displays a relatively complex
collection of lines with notable emissions of [\ion{O}{1}] $\lambda$5577,
[\ion{O}{1}] \dlambda 6300, 6364, [\ion{Ca}{2}] \dlambda 7291, 7324, and
\ion{O}{1} $\lambda$7774.  By day $\sim 158$ the SN has entered the
nebular phase and the spectra show only [\ion{O}{1}] \dlambda 6300, 6364 and
[\ion{Ca}{2}] \dlambda 7291, 7324.  There is a noticeable change in the
emission profile of [\ion{O}{1}] \dlambda 6300, 6364 emission between the two
epochs.  On day 92 the profile shows an asymmetric double-peaked profile with
one peak near 6300 \AA\ and the other blueshifted around 6243~\AA\ ($-2700$
\kms).  However, by day 158 the blueshifted peak has weakened leaving the
[\ion{O}{1}] \dlambda 6300, 6364 emission single-peaked and centered around
6300~\AA.

\subsection{SN 2008bo}

SN 2008bo in NGC~6643 was discovered on 2008 April 22 by \citet{Nissinen08}.
Early optical spectra obtained by \citet{Navas08} showed a close resemblance to
SN~2008ax, and they classified it as a Type Ib SN.  Later spectra
presented here and analyzed with the SuperNova IDentification code (SNID;
\citealt{Blondin07}) favor a Type IIb classification, which we adopt.  Epochs
for SN~2008bo follow from an optical maximum on 2008 April 15 estimated from
unfiltered photometry retrieved from the SNWeb web site.\footnote{Observation
details available at http://www.astrosurf.com/snweb2/2008/08bo/08boMeas.htm.}

Our spectra from days 52, 106, and 162 (Figure 1, {\it right panel}) show
SN~2008bo's entrance into the nebular phase.  Blueshifted [\ion{O}{1}]
$\lambda$5577 emission is observed on days 52 and 106 but faded and was not
detected on day 162.  During this time, emission from the [\ion{O}{1}] \dlambda
6300, 6364 lines shows conspicuous evolution.  On day 52, two peaks, one
centered around 6232 \AA\ and another centered around 6297 \AA\ (approximately
$-3200$ and 0 \kms), are seen, with the blueshifted peak stronger than the
other.  On day 106 the same two peaks are observed, but the blue peak is now
weaker than the red one, and evidence of an additional minor peak around 6268
\AA\ is observed. By day 162, the blue peak continues to weaken in strength
relative to emission centered around $6300$ \AA. In contrast, the [\ion{Ca}{2}]
\dlambda 7291, 7324 lines show little relative change throughout this time
period. 

\begin{figure*} 
\centering 
\includegraphics[width=0.32\linewidth]{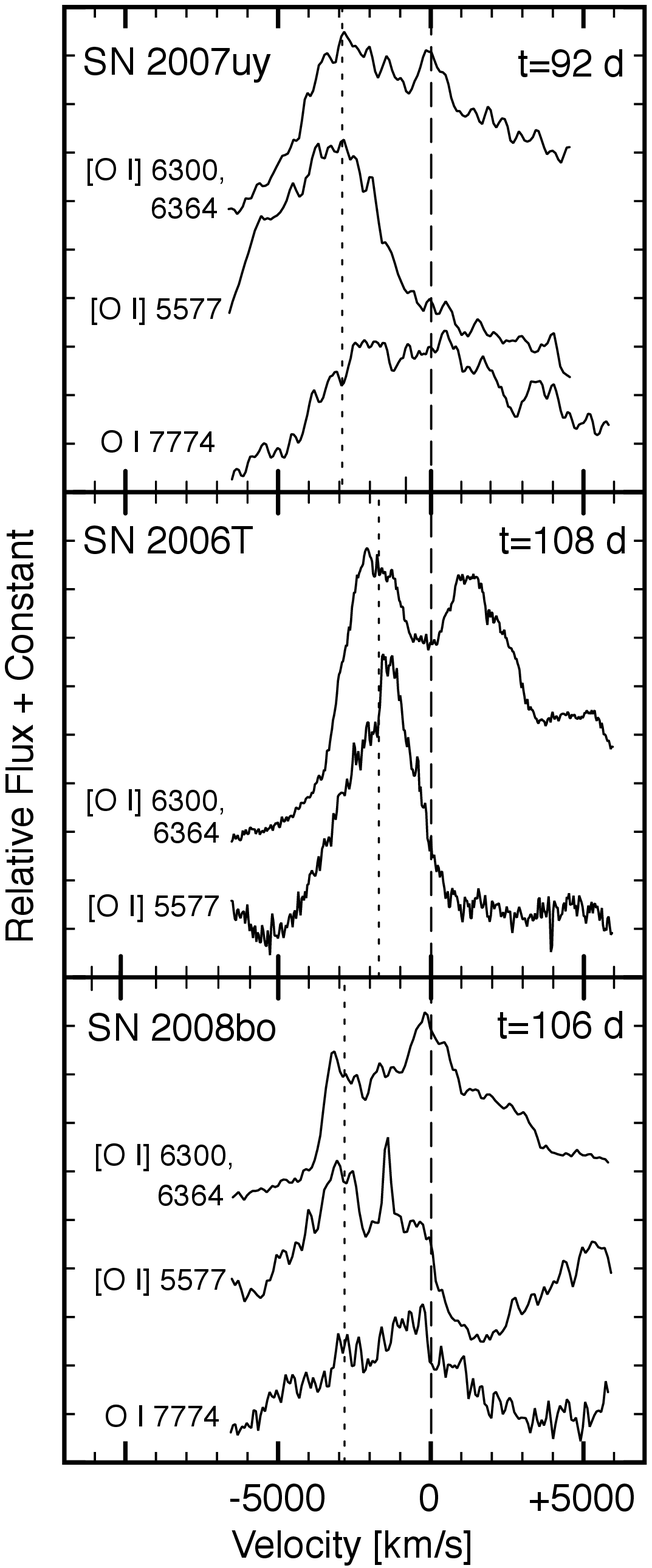}
\includegraphics[width=0.32\linewidth]{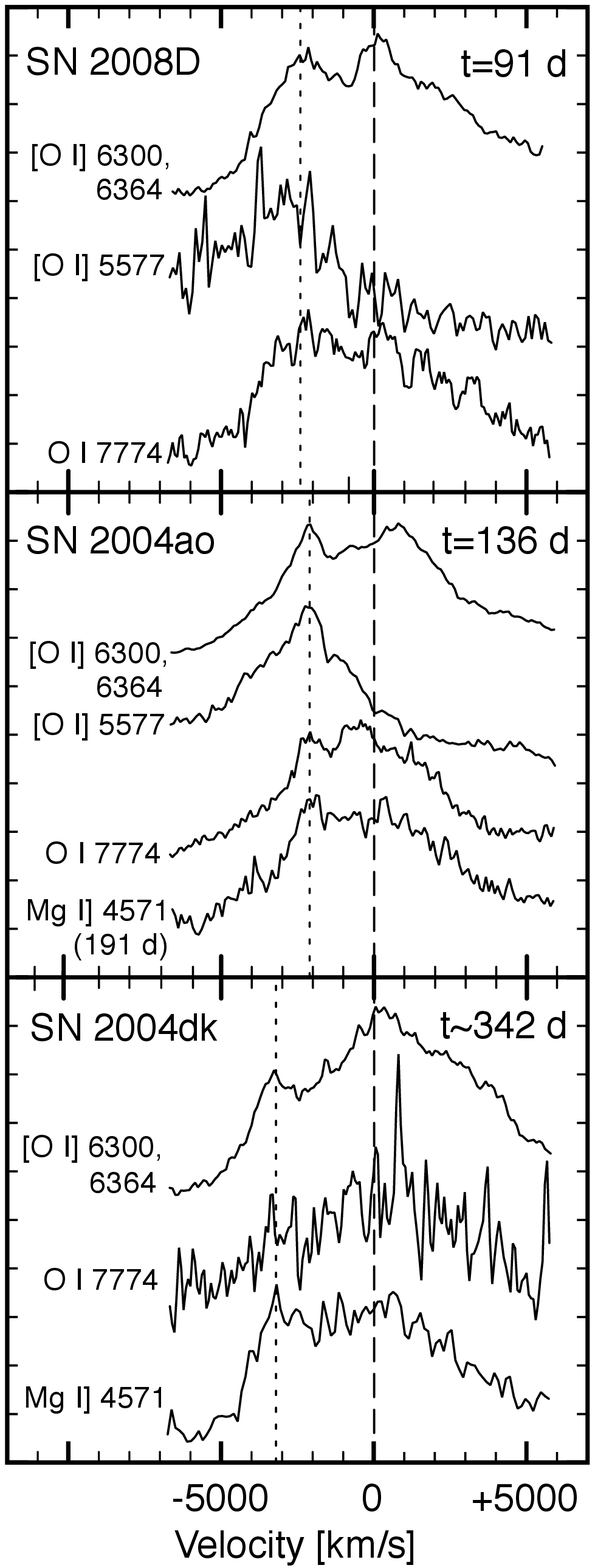} 
\caption{Common blueshifted
peaks in the emission-line profiles of ejecta tracing elements observed in
SNe at nebular epochs of optical evolution.  The heavy dashed line marks
zero velocity with respect to 4571, 5577, 6300, and 7774 \AA\ in the rest frame
of the SNe, while the fainter line highlights conspicuous emission
peaks common across profiles.  Data for SN~1993J are from \citet{Matheson00a}
and \citet{Matheson00b}, SN~2004ao, SN~2004dk, and SN~2006T from
\citet{Modjaz08a}, and SN~2008D from \citet{Modjaz08b}. } 
\end{figure*}

\section{Late-time [\ion{O}{1}] Emission-line Profiles}

Our spectra of five stripped CCSNe show a variety of [\ion{O}{1}] \dlambda
6300, 6364 emission profiles.  SN 2008ax and SN 2007rz show double-peaked
profiles that are basically symmetric about 6300 \AA. SN 2008bo and 2007uy also
show double-peaked profiles but ones exhibiting strong asymmetries at early
epochs.  SN 2007gr, on the other hand, shows principally a single-peaked
profile with a minor asymmetric blueshifted emission peak.  Despite clear
differences between these spectra, some trends emerge in the [\ion{O}{1}]
emission profiles when compared to other SNe.

Below, we discuss in detail the late-time emission-line profiles of SN~2008ax,
which exhibit the sharpest and best-defined [\ion{O}{1}] emission peaks
compared to many other late-time CCSN spectra, and for which we have the best
data set.  Following this discussion, we then compare SN 2008ax's line profiles
against other SNe presented in this paper along with others taken from
the literature.

\subsection{SN~2008ax's O, Ca, and Mg Emission-line Profiles} 

In Figure 4, we show SN~2008ax's [\ion{O}{1}] \dlambda 6300, 6364 line profile
plotted in velocity space relative to other emission-lines at four epochs.  The
top two panels show line profiles for the three neutral oxygen lines
[\ion{O}{1}] \dlambda 6300, 6364, [\ion{O}{1}] $\lambda$5577, and \ion{O}{1}
$\lambda$7774 from spectra obtained on days 74 and 101.  Both the [\ion{O}{1}]
$\lambda$5577 and [\ion{O}{1}] \dlambda 6300, 6364 profiles show common
blueshifted peaks around $-1900$ \kms, highlighted in this figure with the
light dashed line. The \ion{O}{1} $\lambda$7774 profile appears square-topped
with evidence for a fairly abrupt rise in emission strength close to the
velocity of the [\ion{O}{1}] $\lambda$5577 peak and blue [\ion{O}{1}] \dlambda
6300, 6364 peak. 

The bottom two panels show line profiles for days 307 and 394.  Emission
observed at these later times should not suffer as much from radiative transfer
effects or line contamination that may affect the earlier epoch observations of
the above panels.  

Compared to earlier epochs, [\ion{O}{1}] $\lambda$5577 and \ion{O}{1}
$\lambda$7774 emission has faded and is no longer detected.  Blueshifted
\ion{Mg}{1}] $\lambda$4571 emission is observed with a sharp emission peak at a
velocity closely matching the blueshifted peak of the [\ion{O}{1}] \dlambda
6300, 6364 line profile peak of the same epoch and the blueshifted peak of the
[\ion{O}{1}] $\lambda$5577 profile of the two earlier epochs.  Though of 
poorer S/N, our day 394 spectrum shows the same blueshifted peak in the
\ion{Mg}{1}] 4571 line, with possible additional emission around zero velocity.
The profile of [\ion{Ca}{2}] \dlambda 7291, 7324 shows a slightly blueshifted,
broad distribution that remains relatively unchanged at all epochs (see also
Figure 2). 

\begin{figure*}[htp]
\centering
\begin{tabular}{ll}
\includegraphics[width=0.385\linewidth]{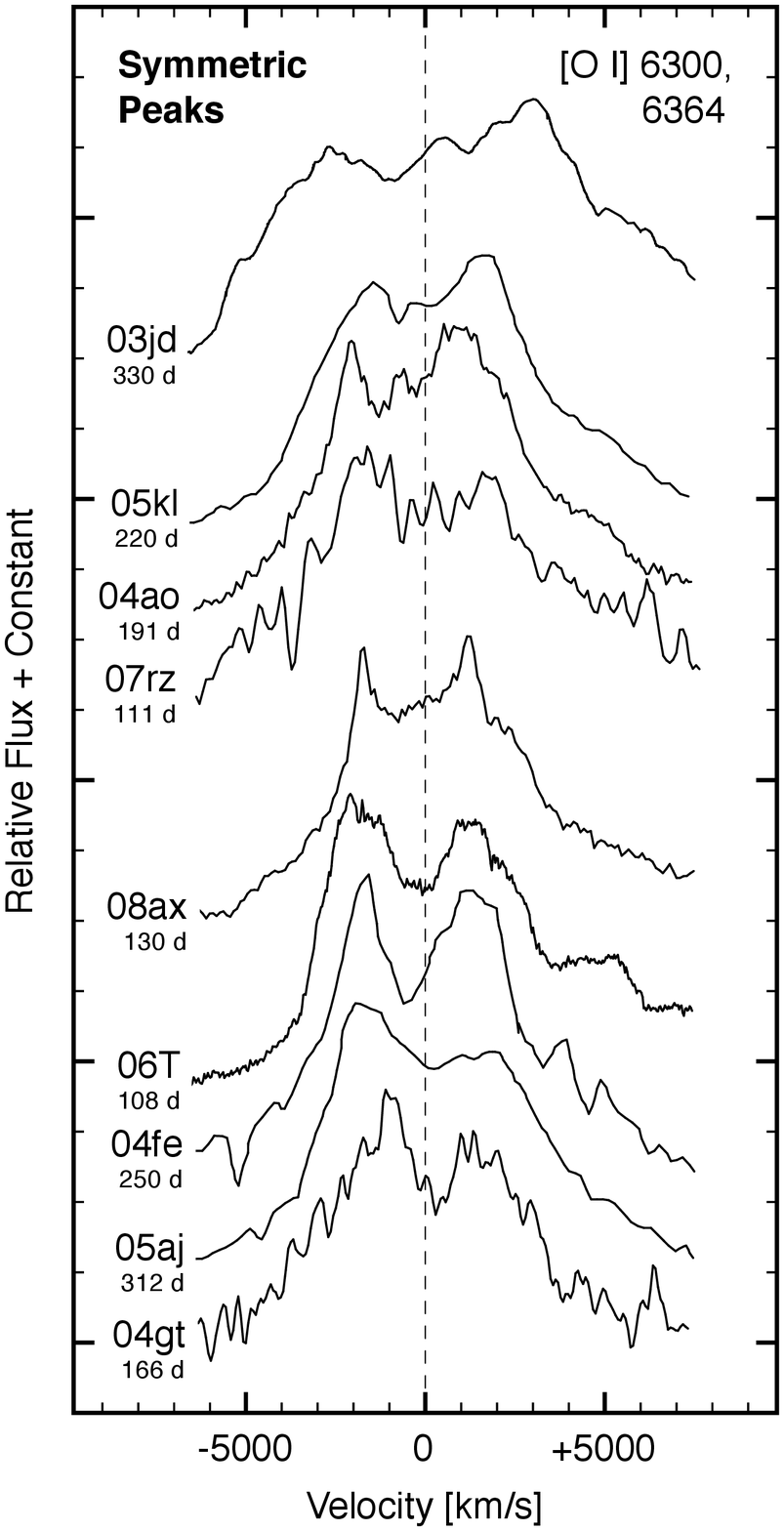} &
\includegraphics[width=0.342\linewidth]{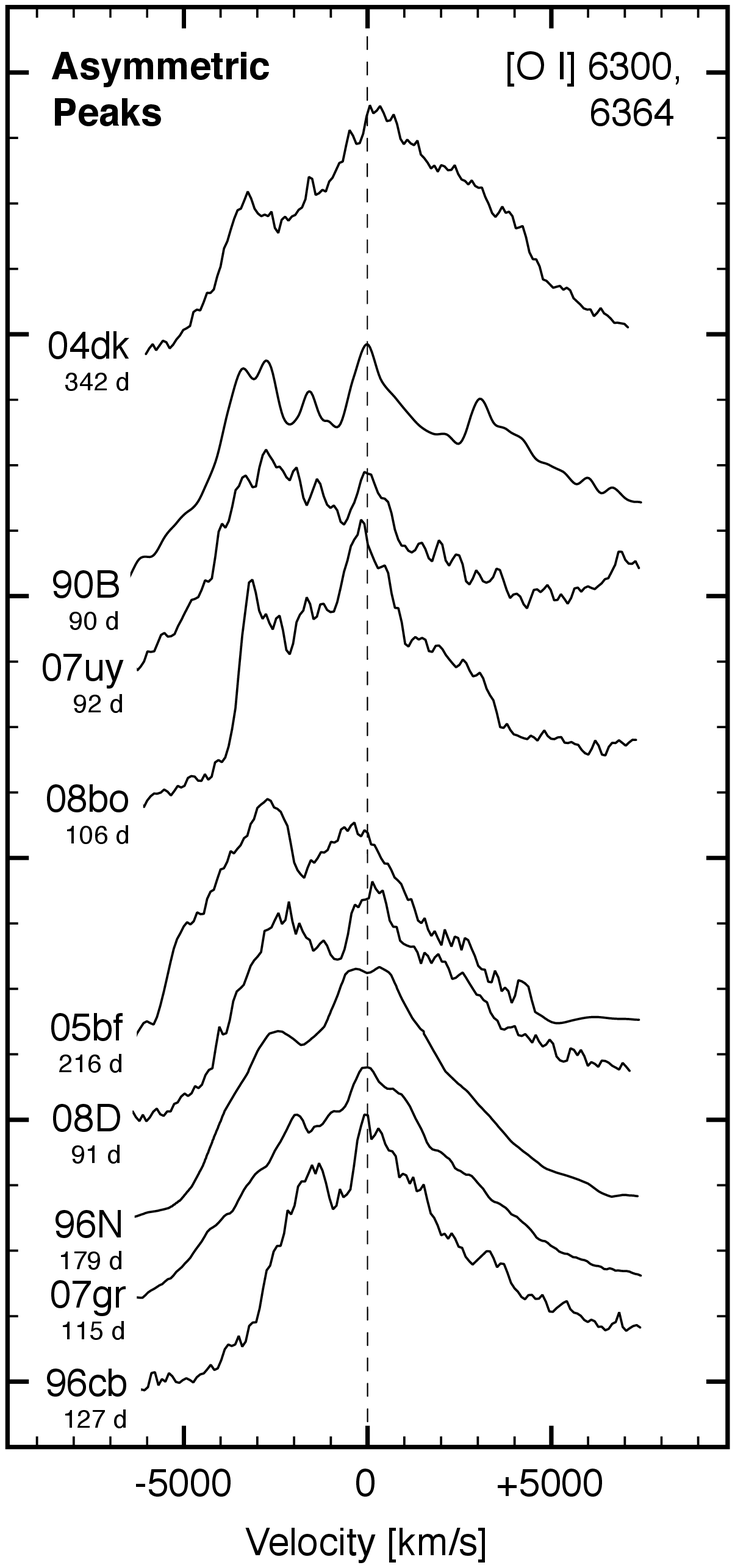}
\end{tabular}
\caption{Classifying [\ion{O}{1}] \dlambda 6300, 6364 emission-line profiles of
CCSNe at nebular epochs of optical evolution.  Velocities are with respect to
6300 \AA\ ({\it dashed line}) in the rest frame of the SNe.  Epochs of
the observations are immediately below abbreviated supernova IDs.  {\it Left
panel}: emission-line profiles approximately symmetric about zero velocity.
{\it Right panel}: asymmetric emission-line profiles skewed toward blueshifted
velocities with one emission peak near zero velocity.} 
\end{figure*} 

In summary, the emission-line profiles of ejecta tracing elements oxygen,
calcium, and magnesium in the late-time spectra of SN~2008ax exhibit quite
different line profiles. [\ion{O}{1}] \dlambda 6300, 6364 shows an unmistakable
double-peaked profile at all epochs observed.  In sharp contrast, the
[\ion{O}{1}] $\lambda$5577 profile shows a prominent single-peaked blueshifted
profile with little redshifted emission, while \ion{O}{1} $\lambda$7774 shows a
broad profile centered close to zero velocity with no obvious emission peaks.
Neither of these lines are observed at the later epochs.  The \ion{Mg}{1}]
$\lambda$4571 profile, which is first observed on day 307, is strongly
blueshifted with an emission peak at a velocity matching the blueshifted peak
seen in the [\ion{O}{1}] \dlambda 6300, 6364 lines. 

\begin{deluxetable*}{lcccccc}
\centering
\tablecaption{References for Figures 6 and 7}
\tablecolumns{7}
\tabletypesize{\footnotesize}
\tablewidth{0pt}
\tablehead{\colhead{Supernova/} &
           \colhead{Host Galaxy} &
           \colhead{Observation} &
           \colhead{Optical Max} &
           \colhead{Epoch} &
           \colhead{Spec.\ Res.} &
           \colhead{Sources/} \\
           \colhead{Type} &
           \colhead{} &
           \colhead{Date} &
           \colhead{Date} &
           \colhead{(days)} &
           \colhead{(\AA)} &
           \colhead{Notes}}
\startdata
SN 1990B~~(Ic)   & NGC 4568 & 1990 Apr 18 & 1990 Jan 19 &
90\tablenotemark{\ }  & 10\tablenotemark{a} & ~1\tablenotemark{b}  \\
SN 1993J~~(IIb)  & NGC 3031 & 1993 Sep 25 & 1993 Mar 27 &
182\tablenotemark{\ }
& 7               & ~2\tablenotemark{c}   \\
SN 1996cb~(IIb)  & NGC 3510 & 1997 May 11 & 1997 Jan 04 &
127\tablenotemark{\ }
& 7 & ~3\tablenotemark{b} \\
SN 1996N~~(Ib)   & NGC 1398 & 1996 Sep 07 & \nodata    &
179\tablenotemark{d}
& \nodata & ~4\tablenotemark{d} \\
SN 2003jd~(Ic-BL) & MCG-01-59-21 & 2004 Sep 12 & 2003 Nov 01 &
330\tablenotemark{\ } & 5 & ~6\tablenotemark{d}  \\
SN 2004ao~(Ib) & UGC
10862 & 2004 Sep 14 & 2004 Mar 07 & 191\tablenotemark{\ } & 10 &
~7 \\
SN 2004dk~~(Ib) & NGC 6118 & 2005 Jul 09 &
\nodata & 342\tablenotemark{e} & 10 & ~7 \\
SN 2004fe~(Ic) & NGC 132  & 2005 Jul 06 & \nodata & 250\tablenotemark{e}
& 10 &
~8\tablenotemark{d}  \\
SN 2004gt (Ic) & NGC 4038 & 2005 Jun 09 &
2004 Dec 24 & 166\tablenotemark{\ } & 10 & ~7 \\
SN 2005aj (Ic) & UGC 02411 & 2005 Dec 27 & \nodata &
312\tablenotemark{e} & 10\tablenotemark{f}
&
~8\tablenotemark{d} \\
SN 2005bf (Ib) & MCG+00-27-5 & 2005 Dec 11 & 2005 May 08 &
216\tablenotemark{\ } & 5\tablenotemark{f} & ~7 \\
SN 2005kl (Ic) & NGC 4369    & 2006 Jun 30 & \nodata    &
220\tablenotemark{e} & 10 & ~8\tablenotemark{d} \\
SN 2006T~~(IIb)& NGC 3054    & 2006 Jun 03 & 2006 Feb 14 &
108\tablenotemark{\ } & 5 & ~7 \\
SN 2008D~~(Ib) & NGC 2770 & 2008 Apr 28 & 2008 Jan 28 &
91\tablenotemark{\ } & 5 & 9 \\
\\
SN 2007gr (Ic) & NGC 1058 & 2007 Dec 21 & 2007 Aug 28 &
115\tablenotemark{\ } & 2 & 10\tablenotemark{\ }\\
SN 2007rz (Ic) & NGC 1590 & 2008 Apr 01 & \nodata    &
111\tablenotemark{e} & 7\tablenotemark{f} & 11\tablenotemark{\ } \\
SN 2007uy (Ib) & NGC 2770 & 2008 Apr 01 & \nodata    &
92\tablenotemark{e}  & 7\tablenotemark{f} & 12\tablenotemark{\ } \\
SN 2008ax (IIb) & NGC 4490& 2008 Jun 06 & 2008 Mar 24 &
74\tablenotemark{\ } & 7 & 13\tablenotemark{\ } \\
SN 2008bo (IIb) & NGC 6643 & 2008 Jul 30 & 2008 Apr 15&
106\tablenotemark{\ } & 7 & 14\tablenotemark{\ }
\enddata
\tablenotetext{\ }{Note: Spectra presented in this paper have been grouped separately
 on the bottom.}
\tablenotetext{a}{Estimated from spectra using narrow H$\alpha$ line.}
\tablenotetext{b}{Downloaded from SUSPECT at
   http://bruford.nhn.ou.edu/$\sim$suspect/index1.html.}
\tablenotetext{c}{Downloaded from
   http://www.noao.edu/noao/staff/matheson/spectra.html.}
\tablenotetext{d}{Digitally traced from published spectra.}
\tablenotetext{e}{Epoch with respect to discovery date.}
\tablenotetext{f}{Presented spectra have been smoothed with a 3 pixel box car function.}
\tablerefs{ (1) \citealt{Clocchiatti01}
            (2) \citealt{Matheson00a}, \citealt{Matheson00b}
            (3) \citealt{Matheson01}, \citealt{Qiu99}
            (4) \citealt{Sollerman98}
            (5) \citealt{Leonard02}, \citealt{Mazzali02}
            (6) \citealt{Mazzali07}, \citealt{Mazzali05}, \citealt{Valenti08b}
            (7) \citealt{Modjaz08a}
            (8) \citealt{Maeda08}
            (9) \citealt{Modjaz08b}
            (10) \citealt{Valenti08a}
            (11) \citealt{Parisky07}, \citealt{Morrell07}
            (12) \citealt{Blondin08}, \citealt{Nakano08}
            (13) \citealt{Pastorello08}
            (14) \citealt{Nissinen08}.}

\end{deluxetable*}

\subsection{SN~2008ax's Line Profiles Compared to Other SN}  

To test if SN 2008ax's oxygen emission profile behavior is exceptional, we
compared its late-time spectra with two of the four other SNe we observed
(Table 1, Figure 1), along with four additional CCSNe taken from the
literature. This sample is comprised of SN~2004ao (\citealt{Modjaz08a}), SN
2004dk (\citealt{Modjaz08a}), SN~2006T (\citealt{Modjaz08a}), SN~2007uy,
SN~2008D (\citealt{Modjaz08b}), and SN~2008bo.  Selection of these six SNe was
based upon the presence of double-peaked [\ion{O}{1}] \dlambda 6300, 6364 line
profiles and the availability of high quality late-time [\ion{O}{1}]
$\lambda$5577, \ion{O}{1} $\lambda$7774, and \ion{Mg}{1}] $\lambda$4571 line
profile data. 

Figure 5 shows the emission-line profiles of these six SNe.  Multiple
epochs are shown for SN 2004ao because the [\ion{O}{1}] $\lambda$5577 and
\ion{Mg}{1}] $\lambda$4571 lines are seen at different stages of optical
evolution.  With the exception of SN 2004dk, these emission profiles observed
at epochs $t < 200$ d may be contaminated by other emission-lines and/or
radiative transfer effects, but are late enough that these effects should be
small.  

Like SN~2008ax, all five [\ion{O}{1}] $\lambda$5577 profiles that are observed
show no appreciable redshifted emission.  Moreover, in four of five of these
cases, but seen particularly clearly in SN~2006T and SN~2004ao, the
[\ion{O}{1}] $\lambda$5577 line profile is single-peaked with the velocity of
the peak closely matching the blueshifted peak in the [\ion{O}{1}] \dlambda
6300, 6364 line profiles.   

Also like SN 2008ax, the \ion{O}{1} $\lambda$7774 and \ion{Mg}{1}] $\lambda
$4571 lines of these SNe exhibit profiles noticeably different from [\ion{O}{1}]
$\lambda$5577.  Most clear is that redshifted emission is not as strong as
blueshifted emission in both lines, and both \ion{O}{1} $\lambda$7774 and
\ion{Mg}{1}] $\lambda $4571 lines sometimes show evidence of blueshifted
emission peaks at velocities matching the blueshifted peaks of the [\ion{O}{1}]
\dlambda 6300, 6364 lines (SN 2008bo, 2008D, 2004ao,
2004dk).  Additional peaks, when observed, are near zero velocity. 

A recent independent survey of [\ion{O}{1}] \dlambda 6300, 6364 line profiles
exhibited in the optical spectra of Ib/c CCSNe by \citet{Taubenberger09}
provides additional examples of \ion{Mg}{1}] $\lambda$4571, [\ion{O}{1}]
$\lambda$5577, and \ion{O}{1} $\lambda$7774 lines to compare.  Many SNe in their
sample show the same trends in the oxygen and magnesium lines.  Examples
include the \ion{Mg}{1}] $\lambda$4571 profile observed in SN 2006T on day 371
(an epoch much later than our data) and in SN 2006ld on day 280, both of which
show single-peaked blueshifted emission at velocities matching blueshifted
peaks in the [\ion{O}{1}] \dlambda 6300, 6364 lines.  Similarly, single-peaked
[\ion{O}{1}] $\lambda$5577 at a blueshifted velocity matching the  \ion{Mg}{1}]
$\lambda$4571 and [\ion{O}{1}] \dlambda 6300, 6364 lines is observed in
SN~2000ew on day 112.   

A caveat to the trends noted above is a possible uncertainty in identifying the
emission feature near 5500~\AA\ with [\ion{O}{1}] $\lambda$5577.  For example,
early analysis of blueshifted emission around 5500~\AA\ in SN~1993J was
identified as [\ion{O}{1}] $\lambda$5577
\citep{Spyromilio94,Filippenko94,Wang94}, but later analysis appeared to favor
[\ion{O}{1}] $\lambda$5577 emission blended with [\ion{Fe}{2}] $\lambda$5536
and [\ion{Co}{2}] $\lambda$5526, the combination of which falsely gave the
impression of blueshifted emission \citep{Houck96}.  However, because of the
correspondence between the blueshifted peaks in the \ion{Mg}{1}] 4571,
[\ion{O}{1}] $\lambda$5577, and [\ion{O}{1}] \dlambda 6300, 6364 lines,
[\ion{O}{1}] $\lambda$5577 appears to be the dominant emission contributor.

\subsection{A Wider Survey of [\ion{O}{1}] \dlambda 6300, 6364 Profiles}

To broaden our investigation of [\ion{O}{1}] \dlambda 6300, 6364 line profiles,
we examined a larger set of stripped-envelope CCSN spectra but this time not
restricted to the availability of high quality late-time [\ion{O}{1}]
$\lambda$5577, \ion{O}{1} $\lambda$7774 or \ion{Mg}{1}] $\lambda$4571 line
profile data.  In Figure 6, we show a comparison of [\ion{O}{1}] \dlambda 6300,
6364 emission-line profiles for a sample of 18 SNe comprised of our five SNe
plus an additional 13 taken from the literature.  Table 2 references the
sources, observing details, and estimated epochs of all spectra.  These 18
objects have been divided into two main types: nine profiles that exhibit two
prominent [\ion{O}{1}] emission peaks positioned approximately symmetrically
about 6300 \AA \ ({\it left panel}), and nine asymmetric profiles with one peak
lying near 6300 \AA \ and additional peaks at blueshifted velocities ({\it
right panel}).

\subsection{Symmetric [\ion{O}{1}] Profiles}

The nine [\ion{O}{1}] \dlambda 6300, 6364 emission-line profiles shown in the
left panel of Figure 6 share the property of exhibiting two conspicuous
emission peaks positioned roughly on either side of 6300 \AA.  The velocity of the
blueshifted emission peak ranges from $\approx -1000$ to $-2600$ \kms\ assuming
it to be due to the [\ion{O}{1}] $\lambda$6300 line.  While the peaks can be
narrow as in the case of 08ax or broad as in 03jd, the presence of these two
emissions peaks dominates the overall appearance of the [\ion{O}{1}] line
profiles.

We subdivided these nine objects into three subgroups.  Sometimes a third,
weaker emission peak near zero velocity is seen in the [\ion{O}{1}] \dlambda
6300, 6364 profile and this is highlighted in the figure for three SNe
05kl, 04ao, and 07rz ({\it left panel, upper middle}).  Contaminating
[\ion{O}{1}] emission from an underlying \ion{H}{2} region is a possible but
unlikely origin for this feature (see \citealt{Tanaka09}).  The broad-lined
Type Ic SN, 03jd, plotted at the top of the figure in its own subgroup,
also shows a weak emission peak near zero velocity.  Such profiles contrast
with five SNe (08ax, 06T, 04fe, 05aj, and 04gt) which exhibit a deep central
trough between peaks with no obvious central emission peak ({\it left panel,
bottom}).  

Although the relative strengths of the blueshifted and redshifted peaks vary
across these subgroups, in most cases the spacing between the two peaks is
close to 3000 km s$^{-1}$ (i.e., $\approx$64 \AA). The two lines of the
[\ion{O}{1}] \dlambda 6300, 6364 doublet (more precisely at 6300.30 and 6363.78
\AA) are suspiciously close to this separation. We measured the center of the
emission peaks for the profiles and found separations between $61-65$ \AA\ in
eight of nine symmetric profiles.

The high incidence of a separation around 64~\AA\ between emission peaks is
illustrated in Figure 7, which is a montage of the eight symmetric peak
profiles sharing this $\approx$64~\AA\ separation. The exception to this group
and not shown in the figure is 03jd, which shows a separation close to twice
this value.  SN 2004gt also appears as an exception in the relatively noisy
spectrum presented here, but the much cleaner day 160 spectrum presented in
\citet{Taubenberger09} confirms a 64 \AA\ separation. 

\begin{figure*}[ht!]
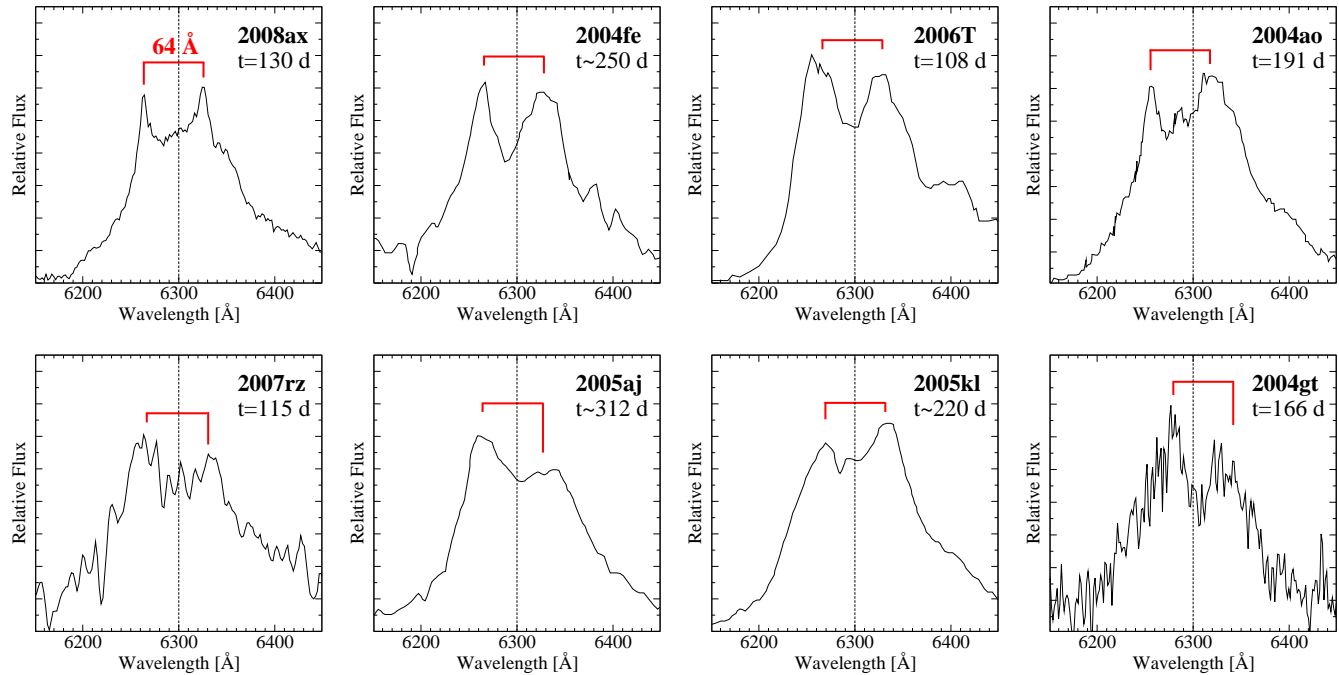
 
\centering \begin{tabular}{cccc}
\includegraphics[width=0.23\linewidth]{f7a.eps} &
\includegraphics[width=0.23\linewidth]{f7b.eps} &
\includegraphics[width=0.23\linewidth]{f7c.eps} &
\includegraphics[width=0.23\linewidth]{f7d.eps} \\ \\
\includegraphics[width=0.23\linewidth]{f7e.eps} &
\includegraphics[width=0.23\linewidth]{f7f.eps} &
\includegraphics[width=0.23\linewidth]{f7g.eps} &
\includegraphics[width=0.23\linewidth]{f7h.eps}
\end{tabular} 
\caption{Eight of the nine late-time [\ion{O}{1}] \dlambda 6300, 6364 emission-line
profiles showing symmetric double peaks from Figure 6.  Separations of 
64~\AA\ (marked in red) between emission peaks are common.} 
\end{figure*}

If the two peaks are associated with the two doublet lines of [\ion{O}{1}]
\dlambda 6300, 6364, the observed 6300:6364 flux ratio of the lines is closer
to 1:1 and not 3:1 as expected from the ratio of their transition probabilities
when the lines are optically thin. This is perhaps unexpected, but not without
precedent, as some SNe have shown similar ratios smaller than the nebular 3:1
value.  Optical spectra of SN~1987A \citep{Spyromilioetal91,Li_n_McCray92} and
1988A \citep{Spyromilio91} showed [\ion{O}{1}] 6300:6364 line ratios close to 1
in spectra taken at epochs $t \sim 200$~d post-outburst.   Though likely
originating from different physical conditions, [\ion{O}{1}] 6300:6364 flux
ratios around 2 and smaller are also routinely observed in novae at similar
times \citep{Williams94}.  Deviations from the 3:1 nebular value are attributed
to optically thick line emission with $\tau \ga 1$ (see \citealt{Li_n_McCray92}
and \citealt{Williams94}).  We investigate other possible interpretations of
these non-nebular ratios in Section $5$.

\subsection{Asymmetric [\ion{O}{1}] Profiles}

The nine [\ion{O}{1}] \dlambda 6300, 6364 emission-line profiles shown in the
right panel of Figure 6 exhibit two or more emission peaks, where one peak is
located close to 6300 \AA, and the other(s) blueshifted with respect to 6300
\AA\ producing an asymmetrical-looking line profile.  As was done for the
symmetric profile SN, we subdivided these asymmetric profiles into three
groups.  SN 2004dk, showing a relatively broad [\ion{O}{1}] profile, forms the
top grouping, while SN showing emission-line profiles that are multi-peaked or
double-peaked form the middle and bottom groupings, respectively. 

Much like that found for many of the symmetric [\ion{O}{1}] profiles shown in
the left panel, a few asymmetric profiles (90B, 07uy, and 08bo; {\it middle
grouping}) exhibit emission peaks separated by $\sim$ 3000 \kms, i.e., close to
the 64~\AA \ separation of the [\ion{O}{1}] \dlambda 6300, 6364 doublet.
However, most show separations between peaks larger or smaller than this ({\it
top and bottom groupings}) with velocities of the blueshifted emission peak
ranging from $-1500$ \kms (SN 1996cb) to $-3300$ \kms (SN 2004dk).  Also unlike
the symmetric profiles plotted in the left panel, most asymmetric profiles show
no conspicuous redshifted emission peaks.  The one apparent exception is 1990B,
although the red emission feature observed around $3000$ \kms\ is likely the
$\lambda$6364 line of [\ion{O}{1}] (see \citealt{Taubenberger09}).

\subsection{The Persistence of Blueshifted Emission Peaks}

An important caveat to Figures 6 and 7 is that the [\ion{O}{1}] \dlambda 6300,
6364 profiles are snapshots of emission profiles at single epochs and thereby
do not address possible evolutionary changes.  From the handful of examples
having multi-epoch observations, symmetric profiles show blueshifted emission
peaks that slowly weaken in strength relative to their redshifted companions.
For example, SN 2008ax, (Figure 2), SN 2004ao \citep{Modjaz07}, and SN 2006T
(compare day 106 of \citealt{Modjaz08a} with day 371 of
\citealt{Taubenberger09}) all exhibit slow but measurable evolution in their
double-peaked [\ion{O}{1}] \dlambda 6300, 6364 line profiles.  Notably, the
emission peaks maintain their blueshifted/redshifted velocities in all cases.

\begin{figure*}[htp!] 
\centering
\includegraphics[width=0.3\linewidth]{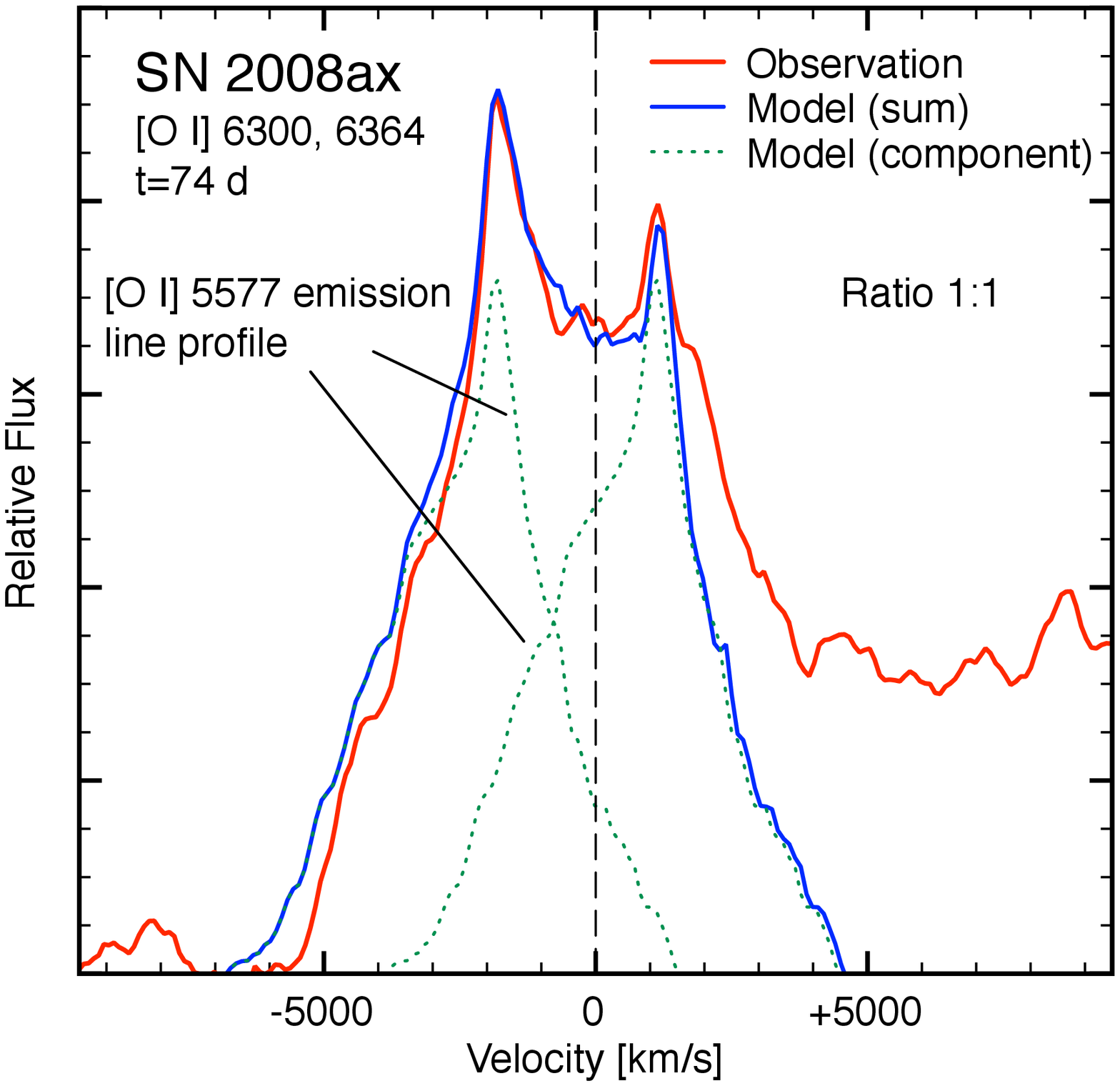}
\includegraphics[width=0.3\linewidth]{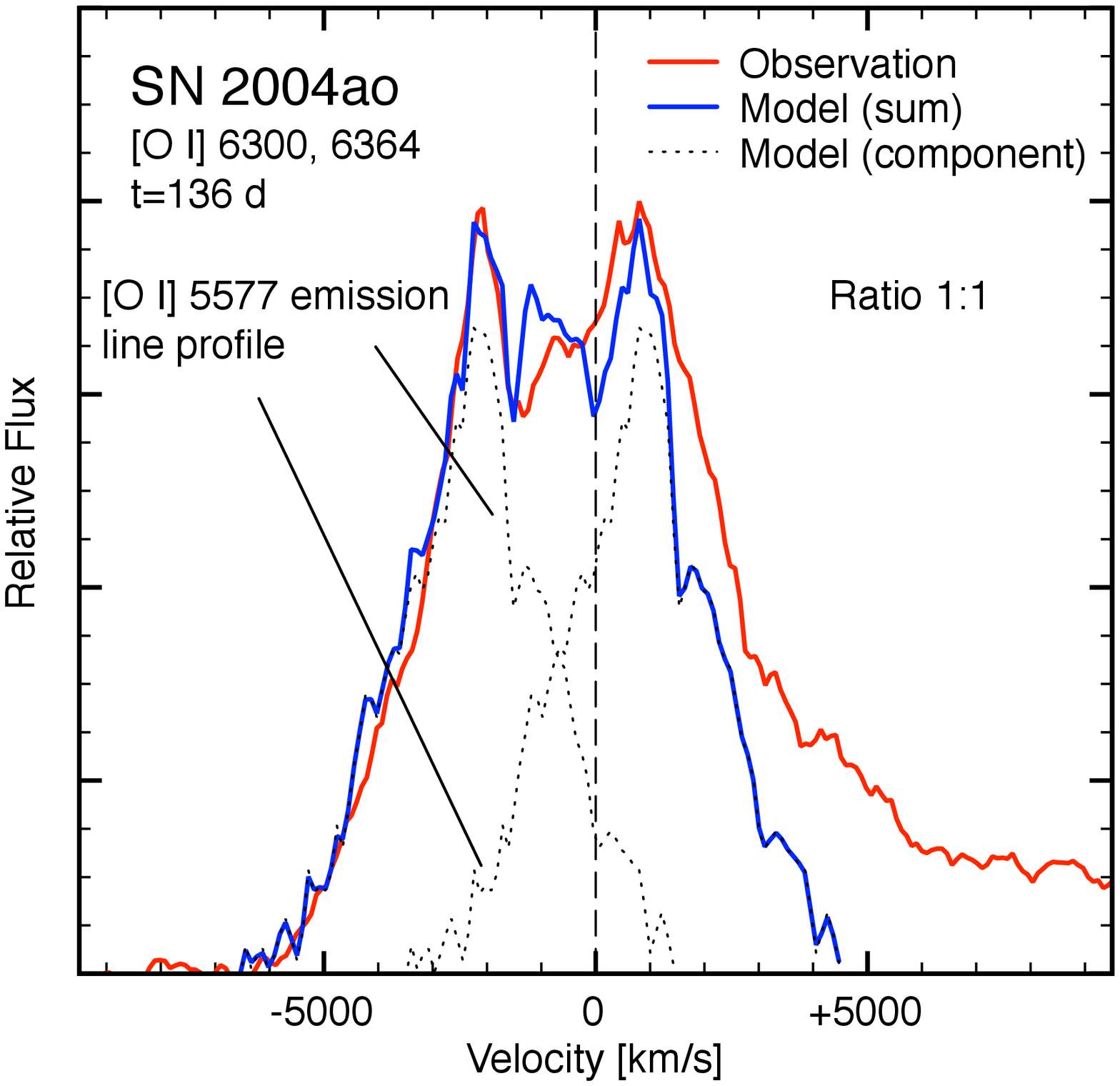}
\includegraphics[width=0.3\linewidth]{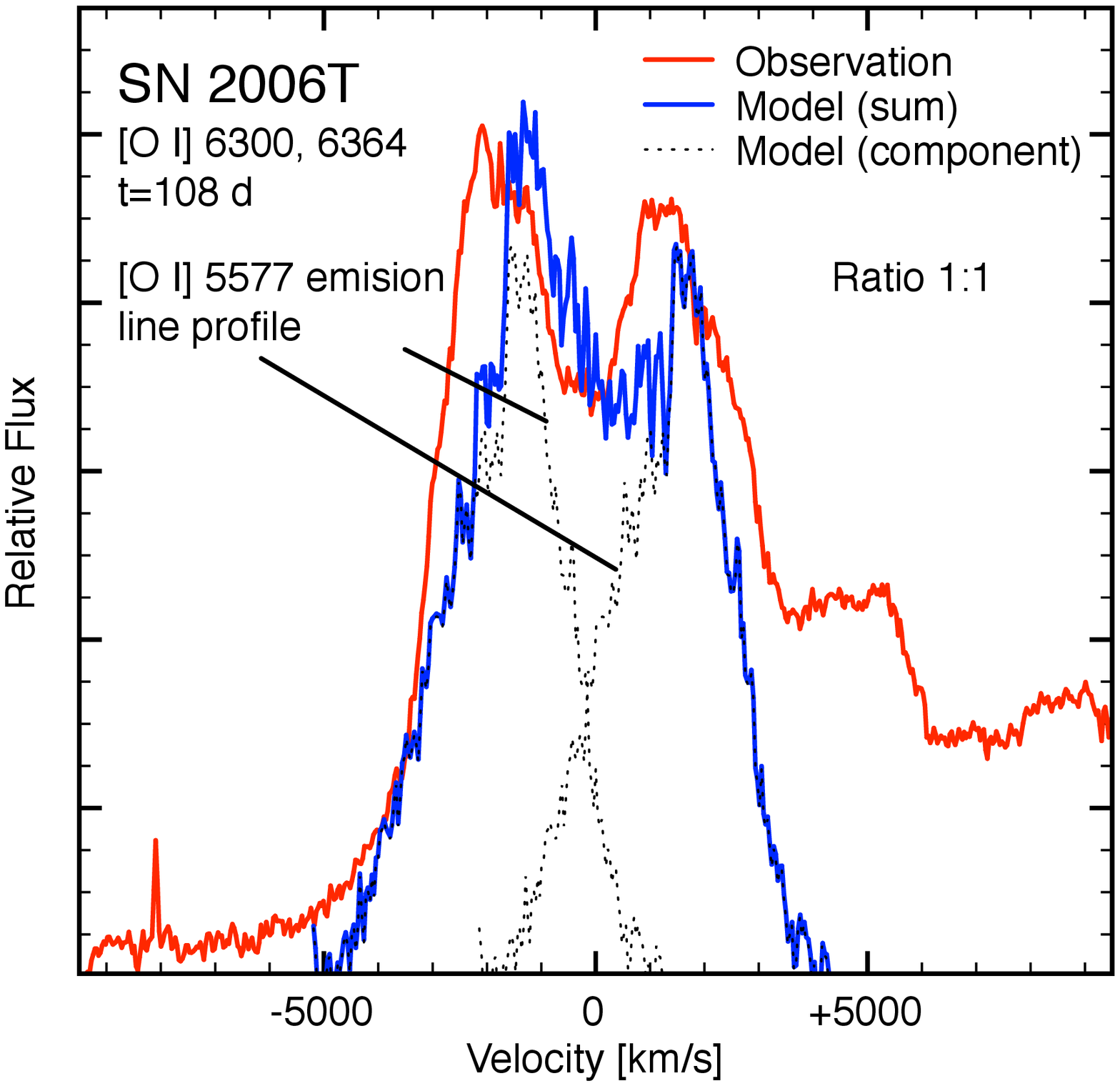} \\
\includegraphics[width=0.3\linewidth]{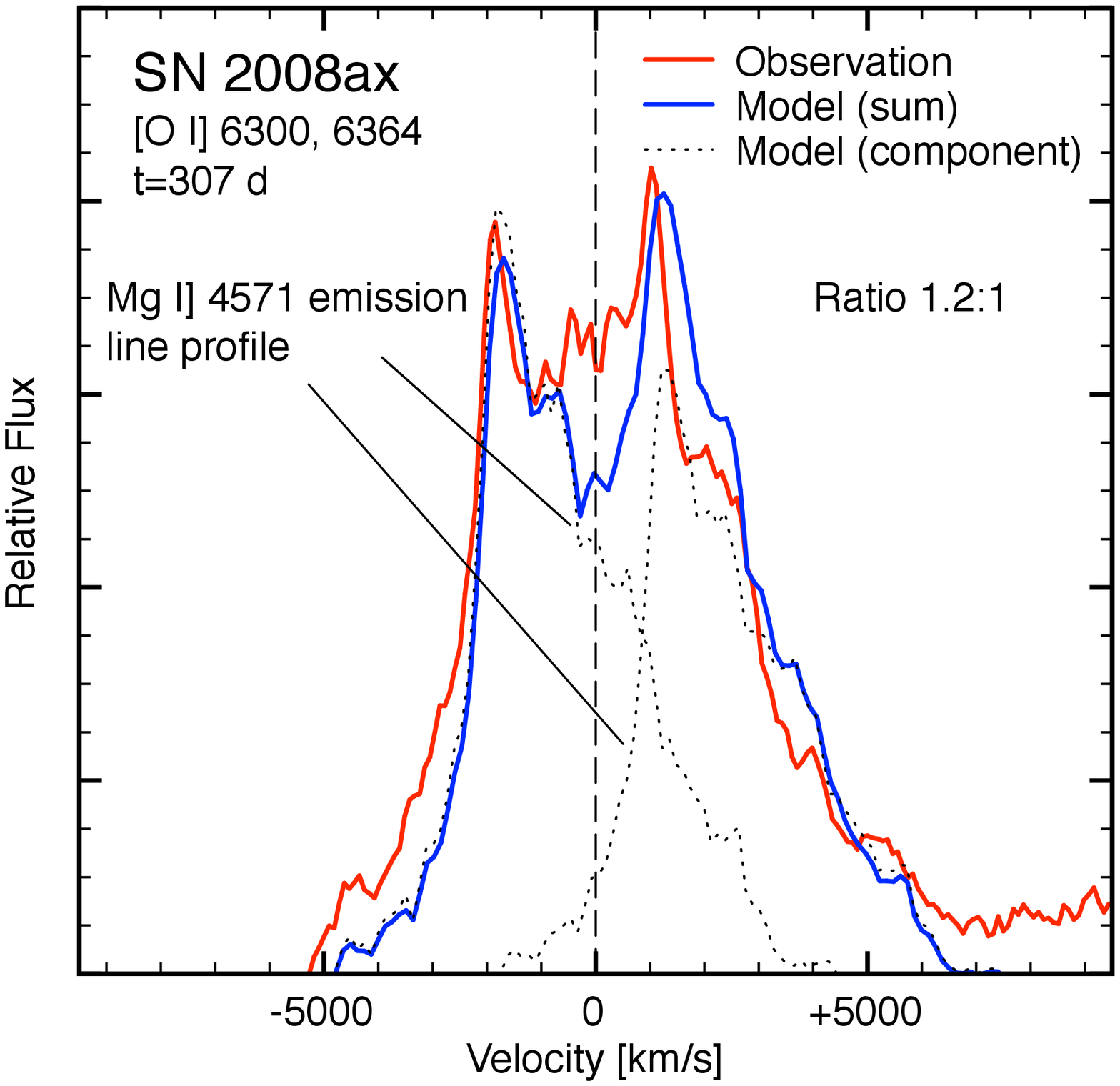}
\includegraphics[width=0.3\linewidth]{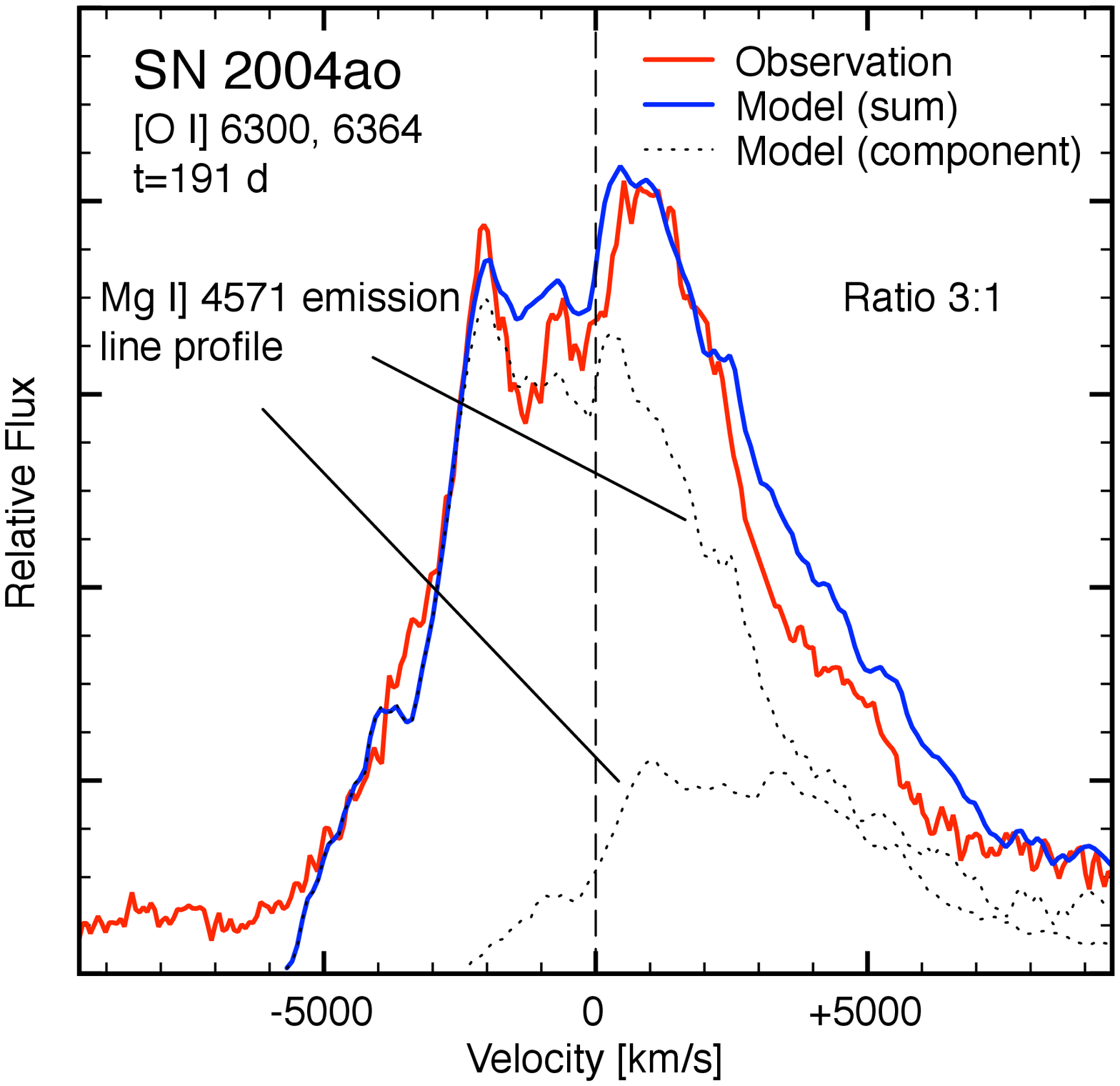}
\includegraphics[width=0.3\linewidth]{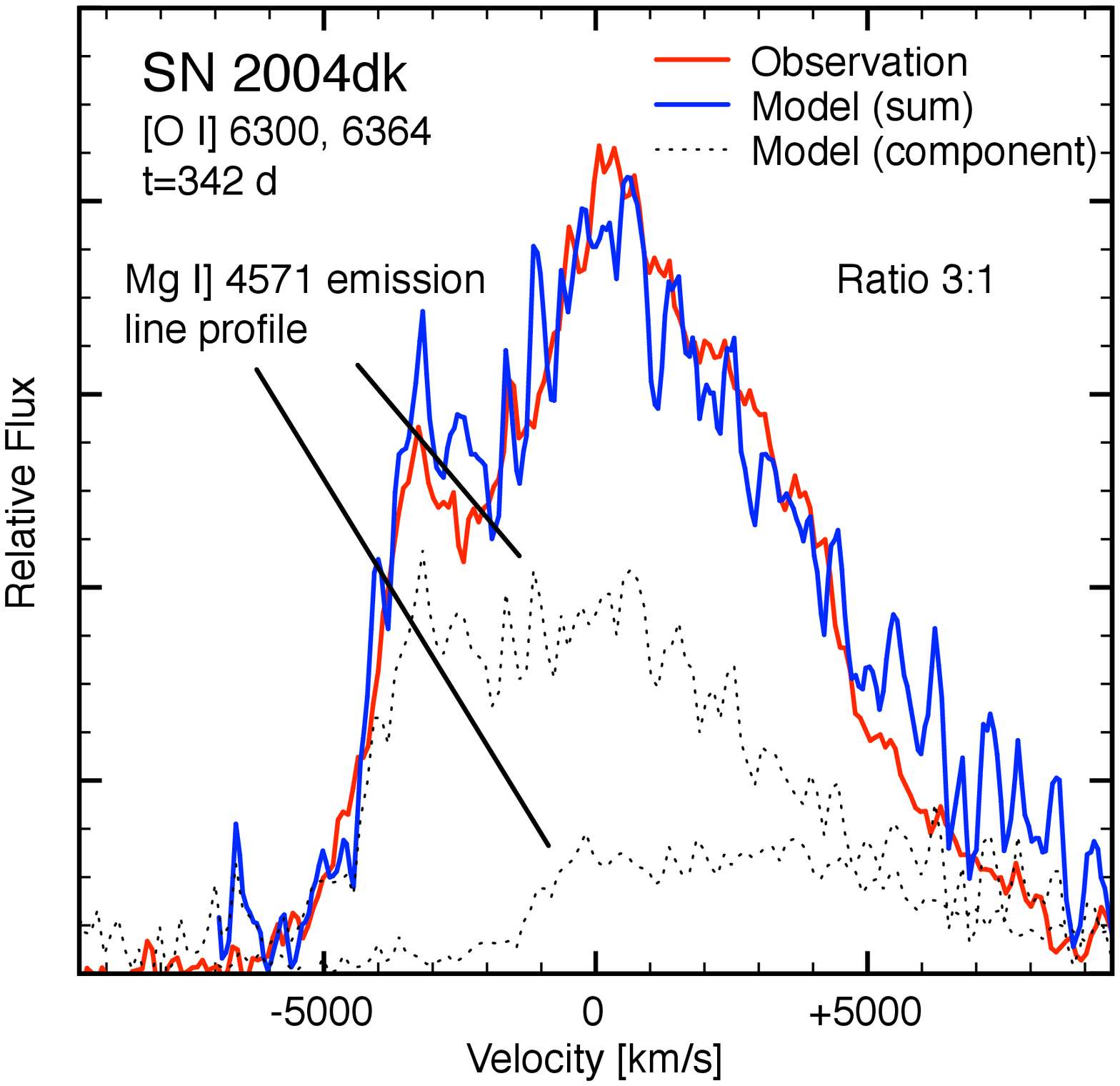}
\caption{Model 1 for late-time [\ion{O}{1}] \dlambda 6300, 6364\
emission: a single preferentially blueshifted emission component. Top panels
are modeled with [\ion{O}{1}] $\lambda$5577 templates, bottom panels with
\ion{Mg}{1}] $\lambda$4571 templates. Observational data for SN~2004ao, 2004dk,
and 2006T are from \citet{Modjaz08a}.} 
\end{figure*}

Asymmetric profiles, on the other hand, show a variety of evolutionary
timescales in their emission peaks.  As demonstrated by the persistent
blueshifted peaks of SN~2004dk and SN~2005bf in Figure 6, and recent
observations of SN~2008D observed on day 363 that show little change from day
91 \citep{Modjaz08b,Tanaka09}, some asymmetric profiles exhibit blueshifted
peaks lasting hundreds of days. However, emission profiles like those seen in
SN~2007uy and SN~2008bo are examples of asymmetric [\ion{O}{1}] \dlambda 6300,
6364 line profiles exhibiting blueshifted emission peaks at early epochs ($t <
90$ d) that diminish in strength at later times.  

Although features observed at epochs $t < 200$ d may be due to contributions
from other emission-lines, the fact that these blueshifted features are
routinely accompanied by blueshifted peaks in other oxygen and magnesium lines
at matching velocities (Figure 5) makes contamination an unlikely source.  As
observed in symmetric profiles, the wavelength of blueshifted emission peaks in
all asymmetric profiles does not change with time, and their strength relative
to emission near zero velocity tends to weaken.

\section{The Nature of [\ion{O}{1}] \dlambda 6300, 6364 Line Profiles}

\subsection{A Torus-like Distribution of O-rich Ejecta?}

Several authors have proposed that double-peaked [\ion{O}{1}] \dlambda 6300,
6364 emission-line profiles are consistent with emission originating from a
torus or disk of O-rich material, possibly orientated perpendicular to a
rapidly expanding jet
\citep{Maeda02,Mazzali05,Maeda06,Maeda08,Modjaz08a,Tanaka09}.  In this model,
an object's double-peaked [\ion{O}{1}] profile reflects the observer's
fortunate orientation close to the plane of an expanding torus/disk of O-rich
ejecta leading to blueshifted and redshifted emission peaks associated with the
approaching and receding portions.  Alternatively, viewing the torus
perpendicular to the plane of expansion would result in a sharp, single-peaked
[\ion{O}{1}] line profile.

Some of the observational trends noted above in Section $4$, however, signal
warnings about adopting a torus/disk interpretation for some double-peaked
[\ion{O}{1}] \dlambda 6300, 6364 profiles.  Most apparent is that the
separation of the two prominent peaks in symmetric [\ion{O}{1}] profiles is
often very close to 64 \AA, i.e., the separation of the [\ion{O}{1}] \dlambda
6300, 6364 doublet (Figure 7).   Instead of the blueshifted and redshifted
peaks representing the forward and rear portions of an expanding torus of
O-rich material, the separation suggests that the two peaks actually originate
from the two lines of the doublet from a single emitting source on the front of
the SN moving toward the observer.

Double emission peaks seen in asymmetric profiles with separations larger or
smaller than the doublet spacing (e.g., SN 2004dk and 2005bf) do not share this
problem.  However, the preferentially blueshifted emission profiles of Figure 6
illustrate the point that asymmetric profiles rarely show prominent peaks at
wavelengths significantly longward of 6300 \AA.   If the two observed emission
peaks are attributed solely to the 6300 \AA\ line of [\ion{O}{1}] originating
from a toroidal geometry of ejecta, then one must interpret the tori
to have centers of expansion having velocities blueshifted
toward the observer, with the most redshifted portion of the expanding
ring/torus at a velocity close to zero.
 
An example of a double-peaked [\ion{O}{1}] \dlambda 6300, 6364 profile
interpreted this way was SN 2005bf (refer to Figure 6).  \citet{Maeda07} modeled
the blueshifted profile as emission originating from a layered blob of Fe,Ca,O
material either (1) unipolar and moving at a center-of-mass velocity $v \sim
2000 - 5000$ \kms, or (2) suffering from self-absorption within the ejecta.  In
this case, the blueshifted double-peaked [\ion{O}{1}] \dlambda 6300, 6364 line
profile was thought to be unique, and thus explanation (1) involving an
extremely elongated shell of O-rich material, though unusual, was not viewed as
improbable especially in light of good model agreement.  

However, Figure 6 suggests that asymmetric profiles like SN 2005bf are not
unique. While blobs of ejecta moving at thousands of \kms\ may represent one
possible scenario (e.g., unipolar explosion models; see
\citealt{Hungerford05}), the frequent need for the centers of expansion of
these blobs to be at velocities systematically directed toward the observer
seems problematic.

\subsection{Other Models}

Complications with a torus/disk model raised by our observations prompted us to
investigate two alternative interpretations of the double-peaked [\ion{O}{1}]
\dlambda 6300, 6364 emission-line profiles.  In light of mounting evidence that
CCSNe are intrinsically aspherical, a torus geometry might still be correct but
with considerable internal extinction.  In cases like SN~2003jd where the
expansion velocity is especially high and extinction is minimal, one may be
truly seeing both blue and red sides of a shell or torus \citep{Mazzali05}.
However, in other cases where velocities are smaller and extinction is higher,
the rear side of the ejecta and its emission might be hidden, even at epochs $t
> 200$ d. 

On the other hand, different line profiles observed in the oxygen and magnesium
lines noted in Section $4$ may originate from different regions of the SN.
Line formation differences attributable to density, temperature, and energy
potential dependencies may be introducing pronounced emission discrepancies
between elements and species.  Hence, the blueshifted, narrow profiles observed
in [\ion{O}{1}] $\lambda$5577 and the zero velocity, broad profiles observed in
\ion{O}{1} $\lambda$7774 may originate from two different regions of the
SN: (1) a central, pseudo-spherically symmetric distribution of O-rich
ejecta, and (2) a clump or shell of O-rich material traveling at a moderate
velocity ($-2000$ to $-4000$ \kms) in the front-facing hemisphere. 

We explored both of these scenarios in simple line fitting models.  Below, we
briefly discuss the results of these models, comment on their interpretations,
and highlight their own associated difficulties.

\begin{figure*}[htp!] 
\centering
\includegraphics[width=0.3\linewidth]{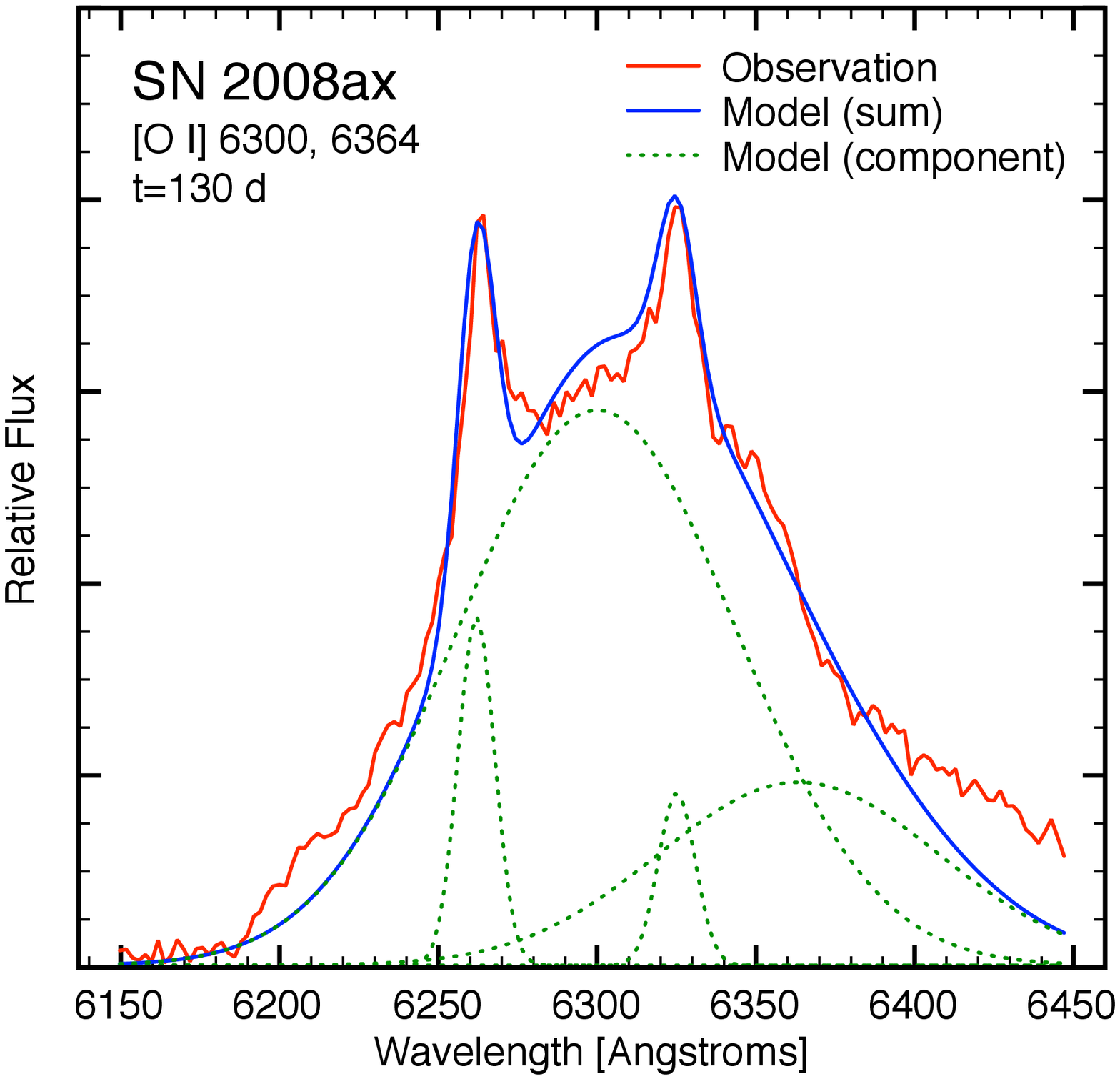}
\includegraphics[width=0.3\linewidth]{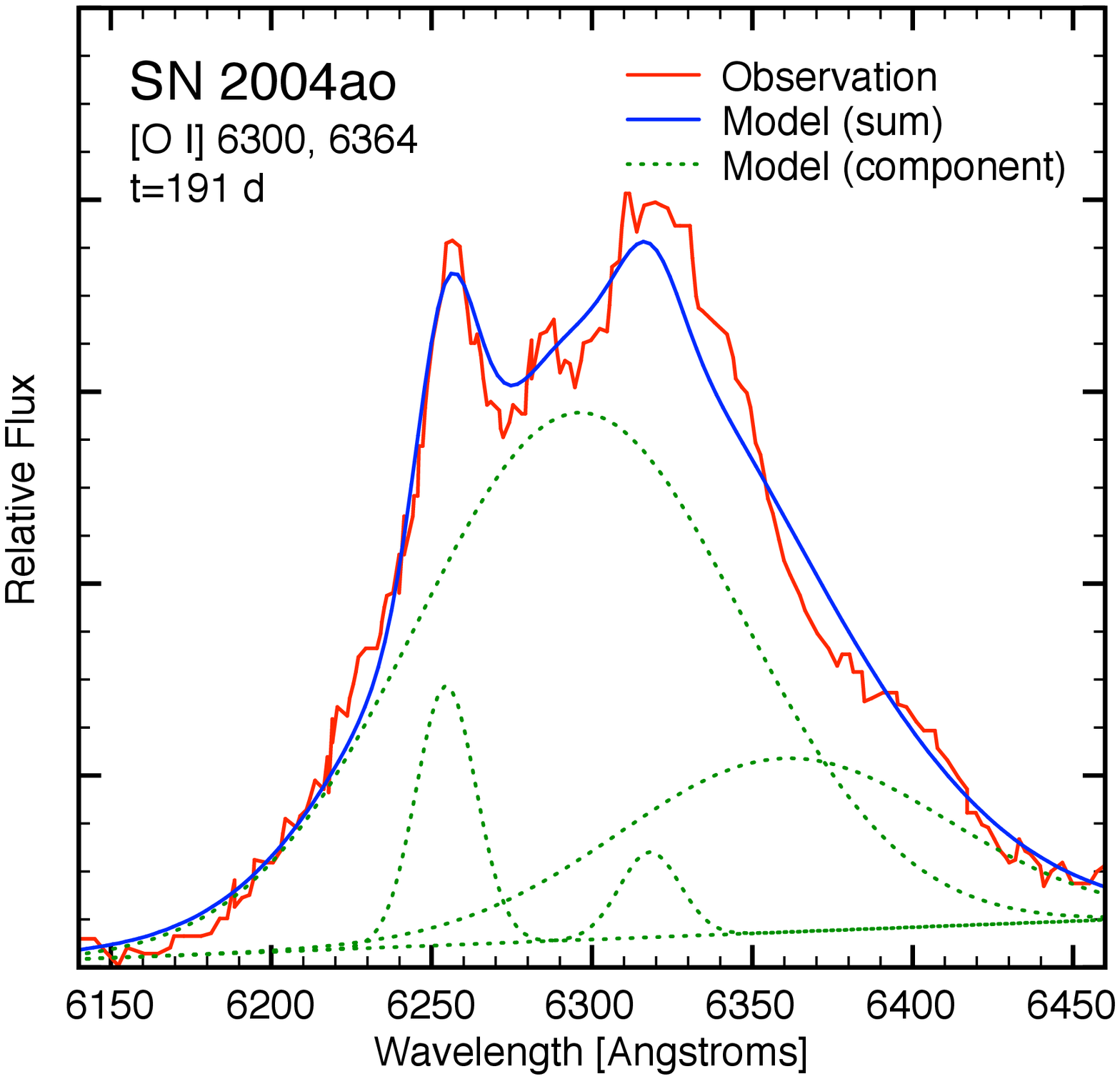}
\includegraphics[width=0.3\linewidth]{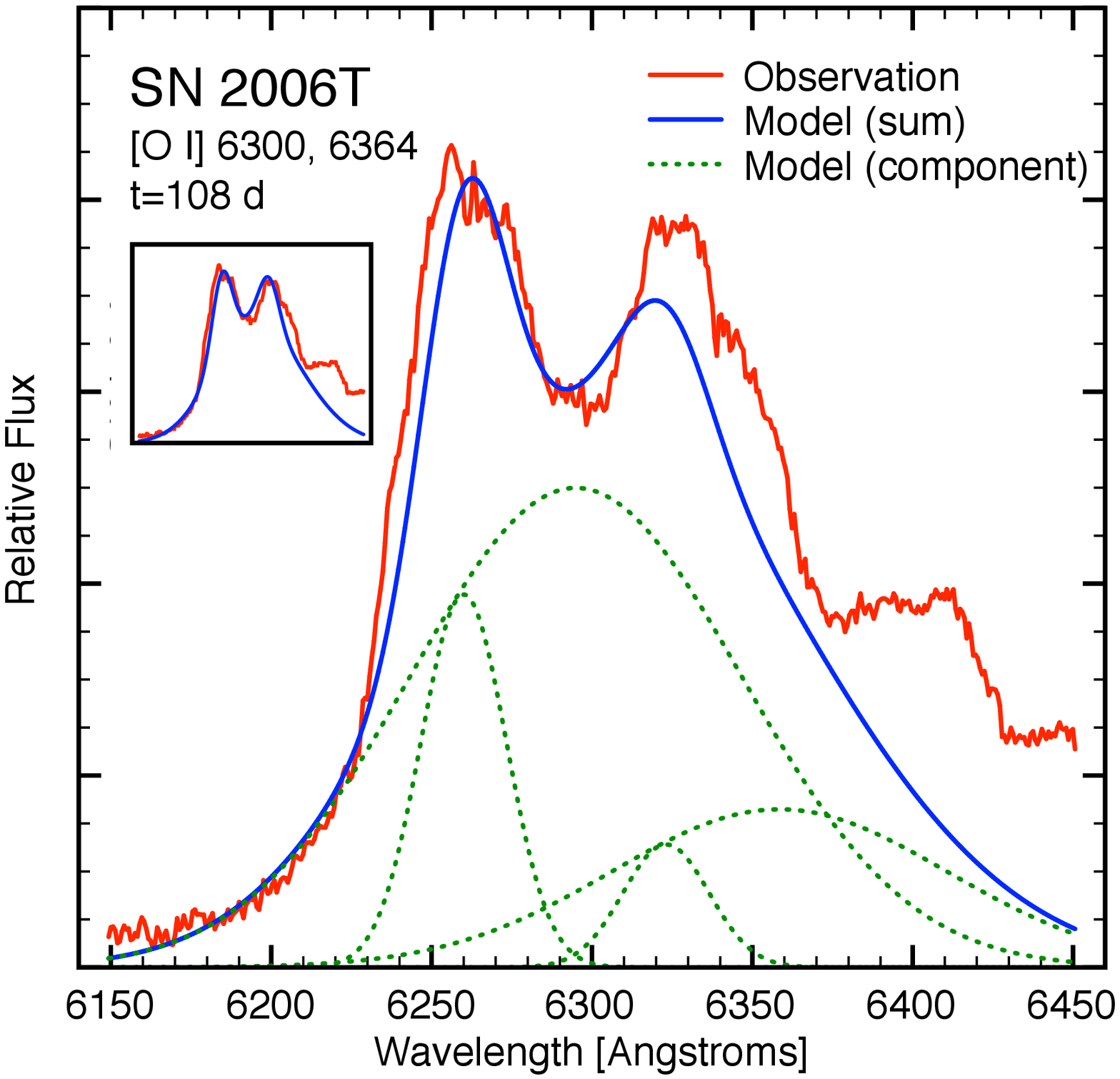} \\
\includegraphics[width=0.3\linewidth]{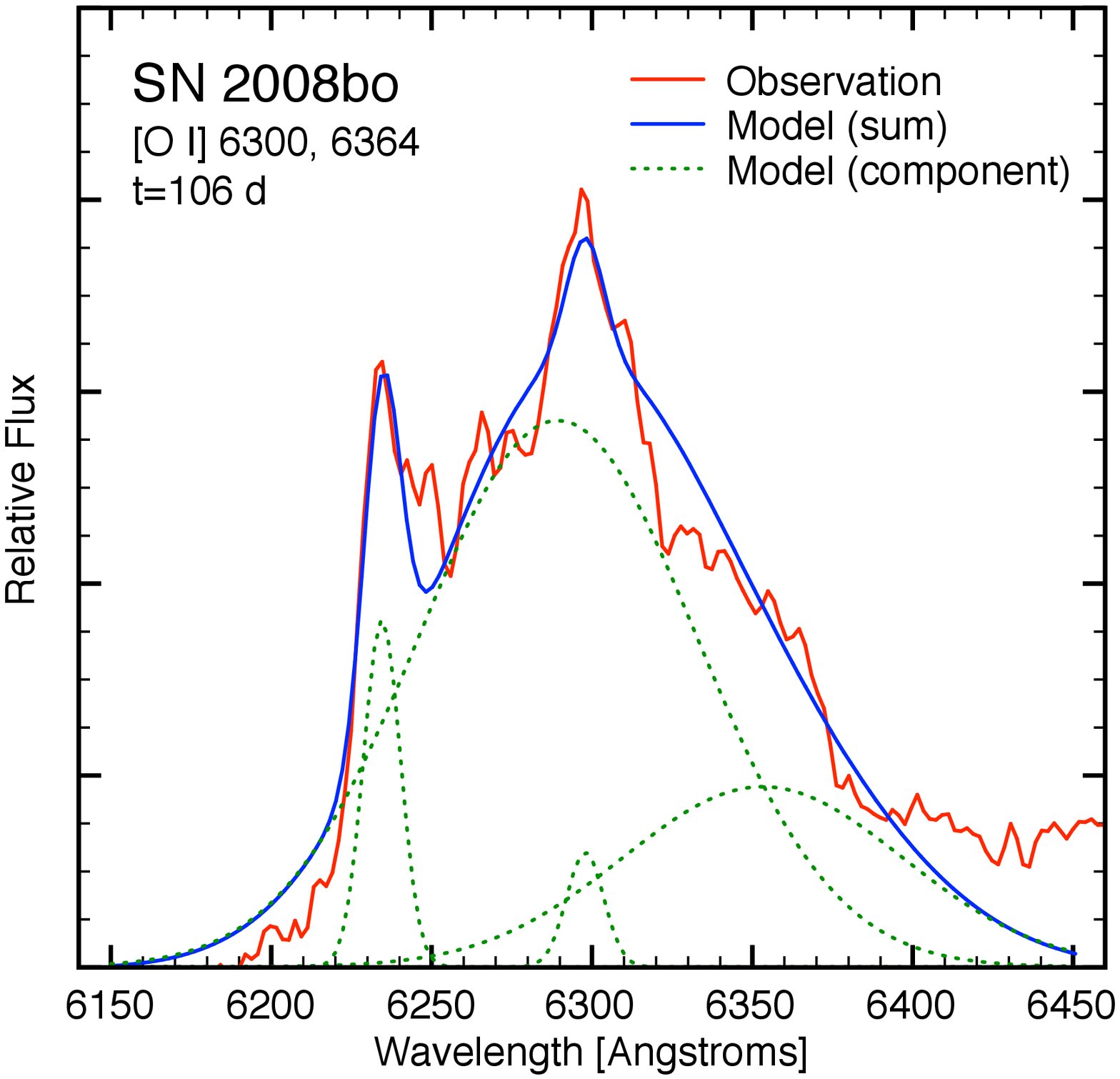}
\includegraphics[width=0.3\linewidth]{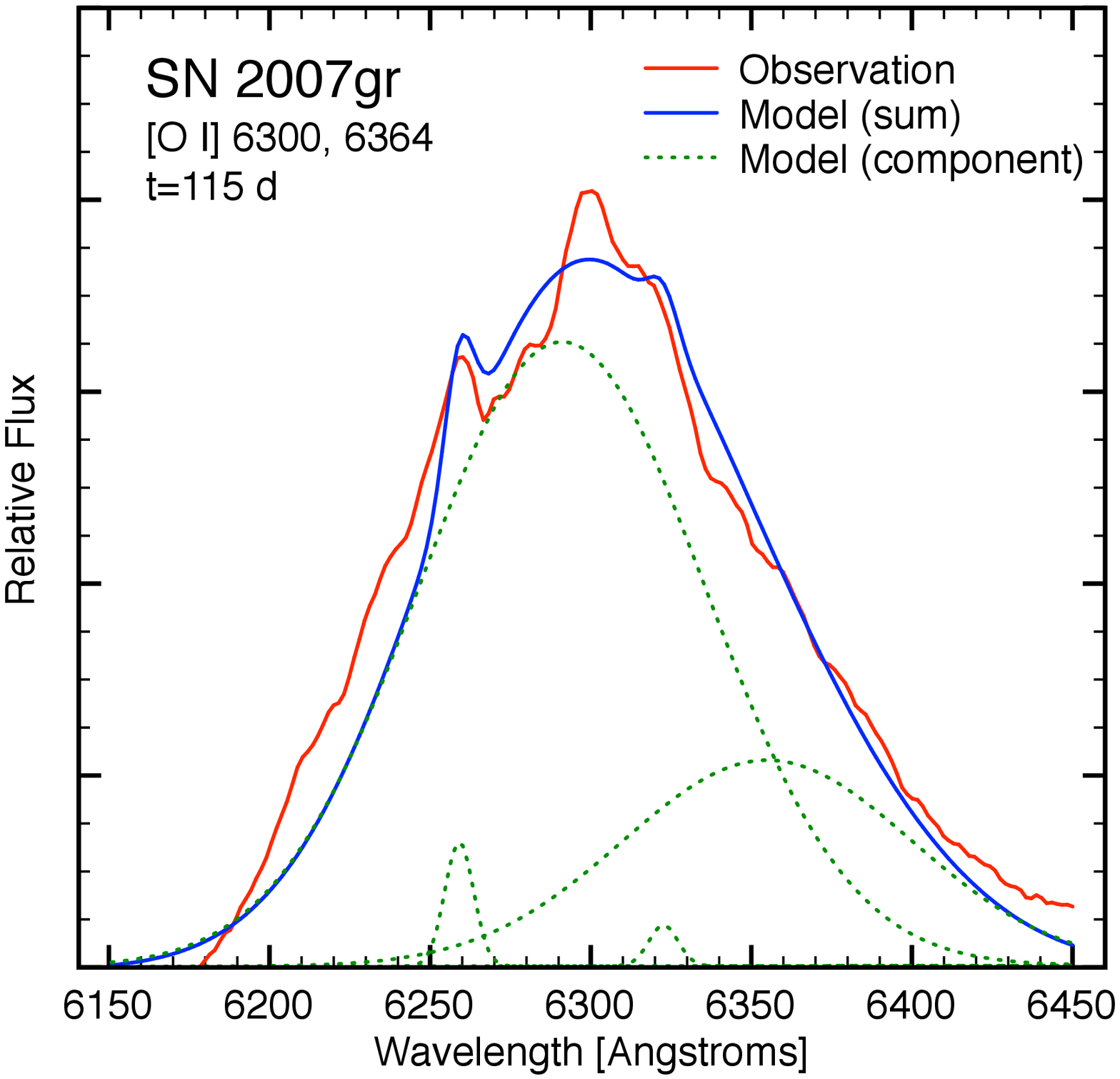}
\includegraphics[width=0.3\linewidth]{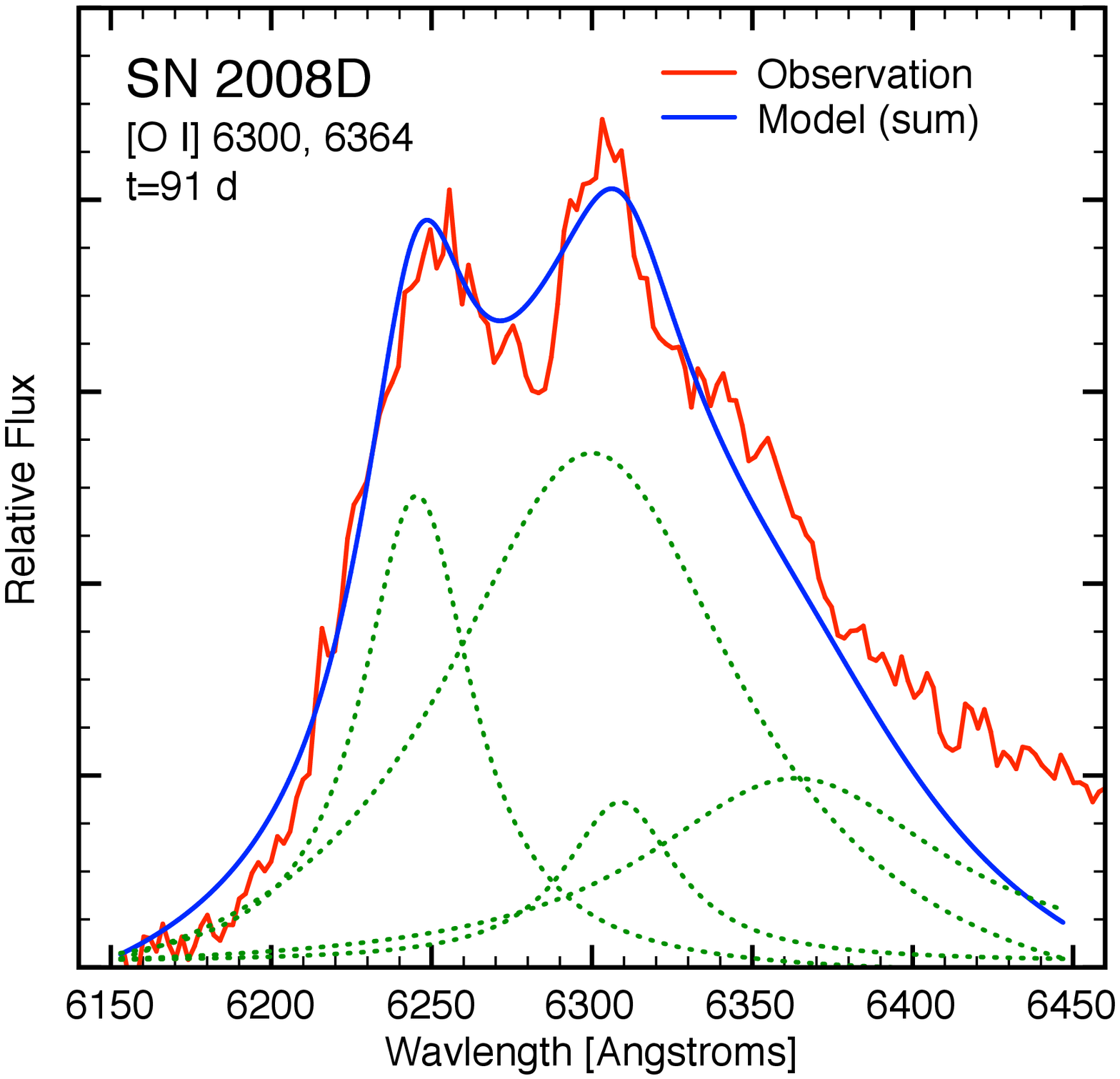}
\caption{Model 2 for late-time [\ion{O}{1}] \dlambda 6300, 6364
emission: broad emission centered around zero velocity with a narrow blueshifted
emission component.  {\it Top}: symmetric profiles.  {\it Bottom}: asymmetric
profiles. Observational data for SN~2004ao are from \citet{Modjaz08a}. All 
line models adopt an [\ion{O}{1}] 6300:6364 flux ratio of 3:1, except for the 
inset of SN 2006T where a ratio of 1.8:1 was used.}
\end{figure*}

\begin{figure*} 
\centering 
\includegraphics[width=0.4\linewidth]{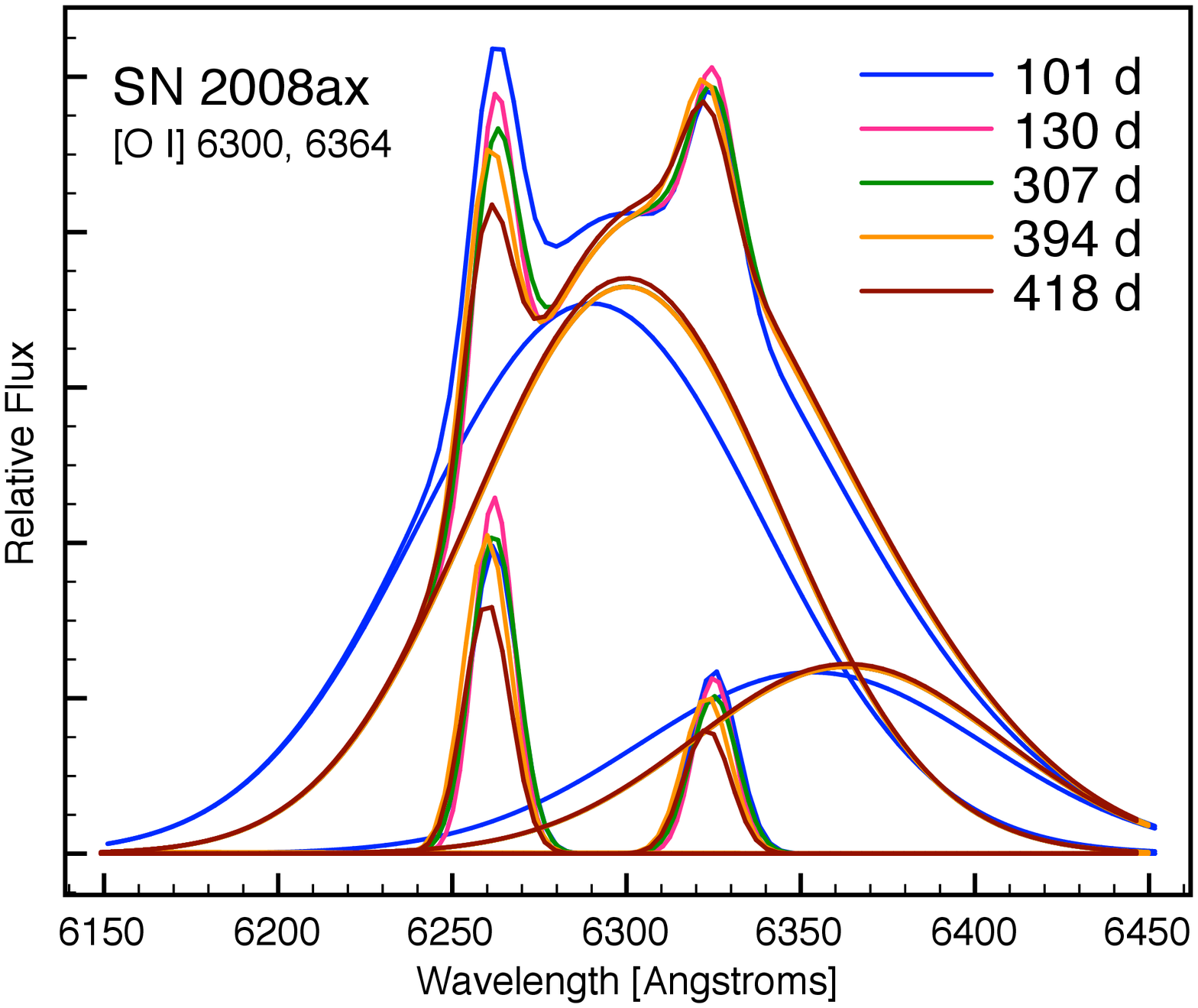}
\includegraphics[width=0.4\linewidth]{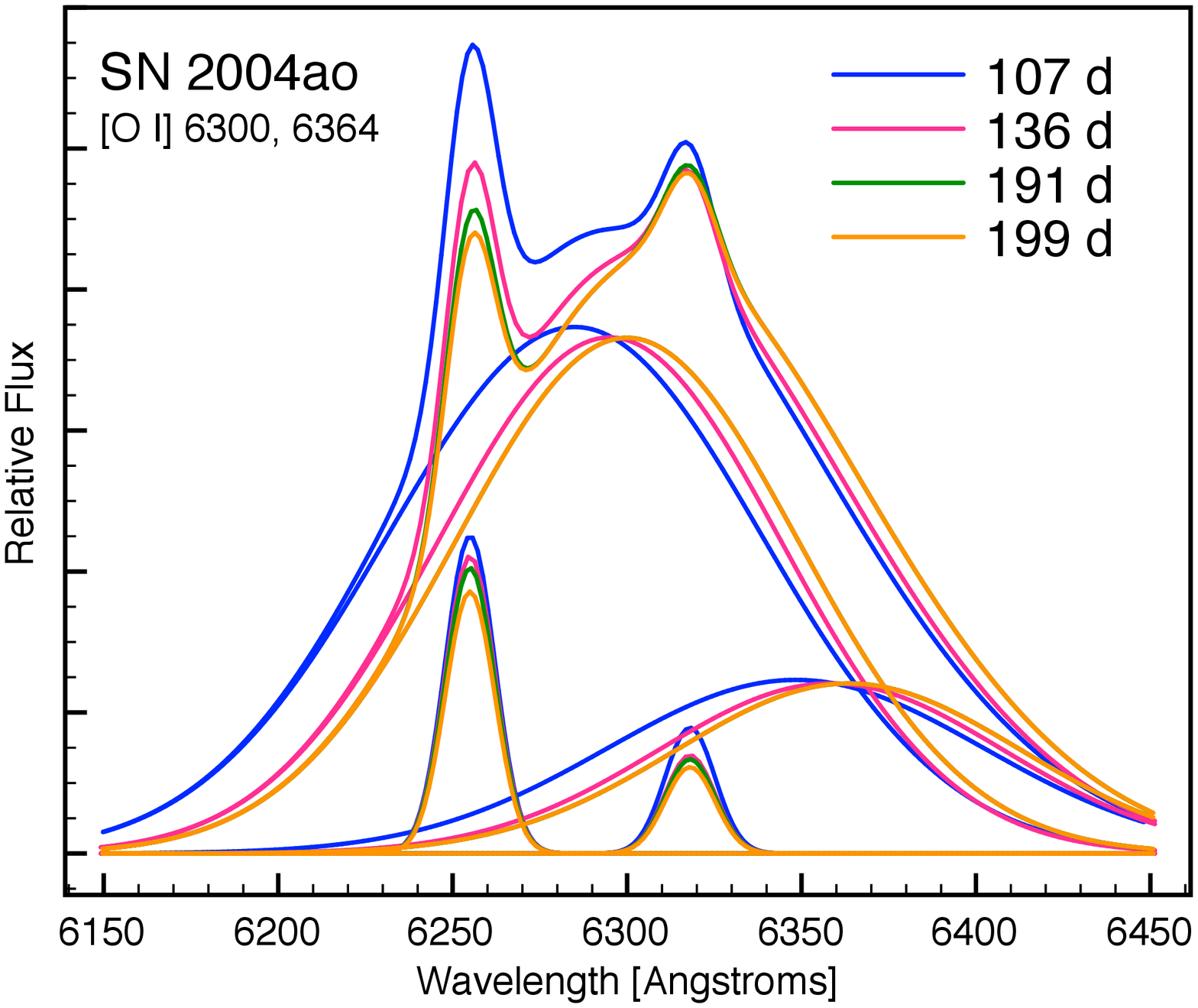} \\
\includegraphics[width=0.4\linewidth]{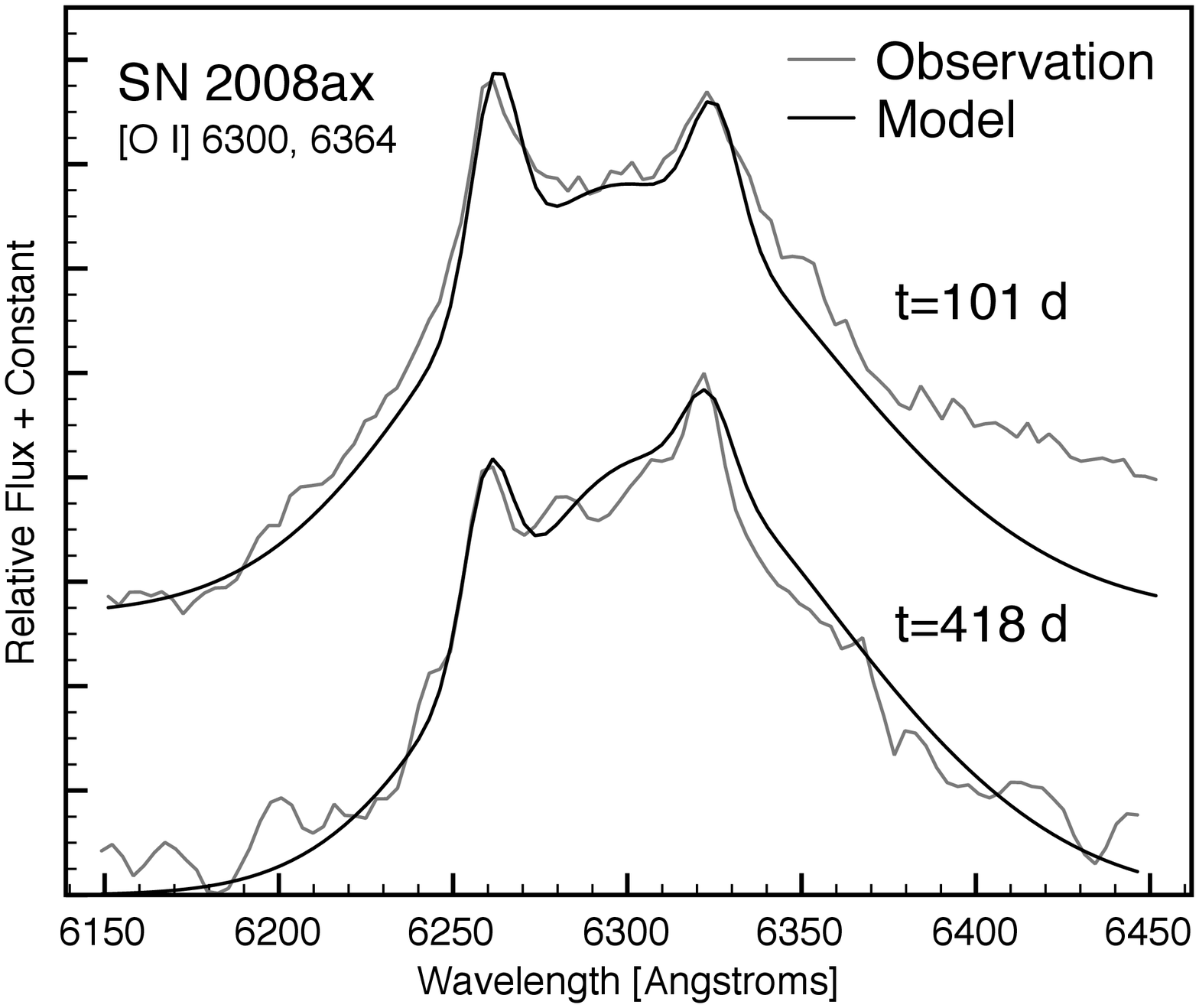}
\includegraphics[width=0.4\linewidth]{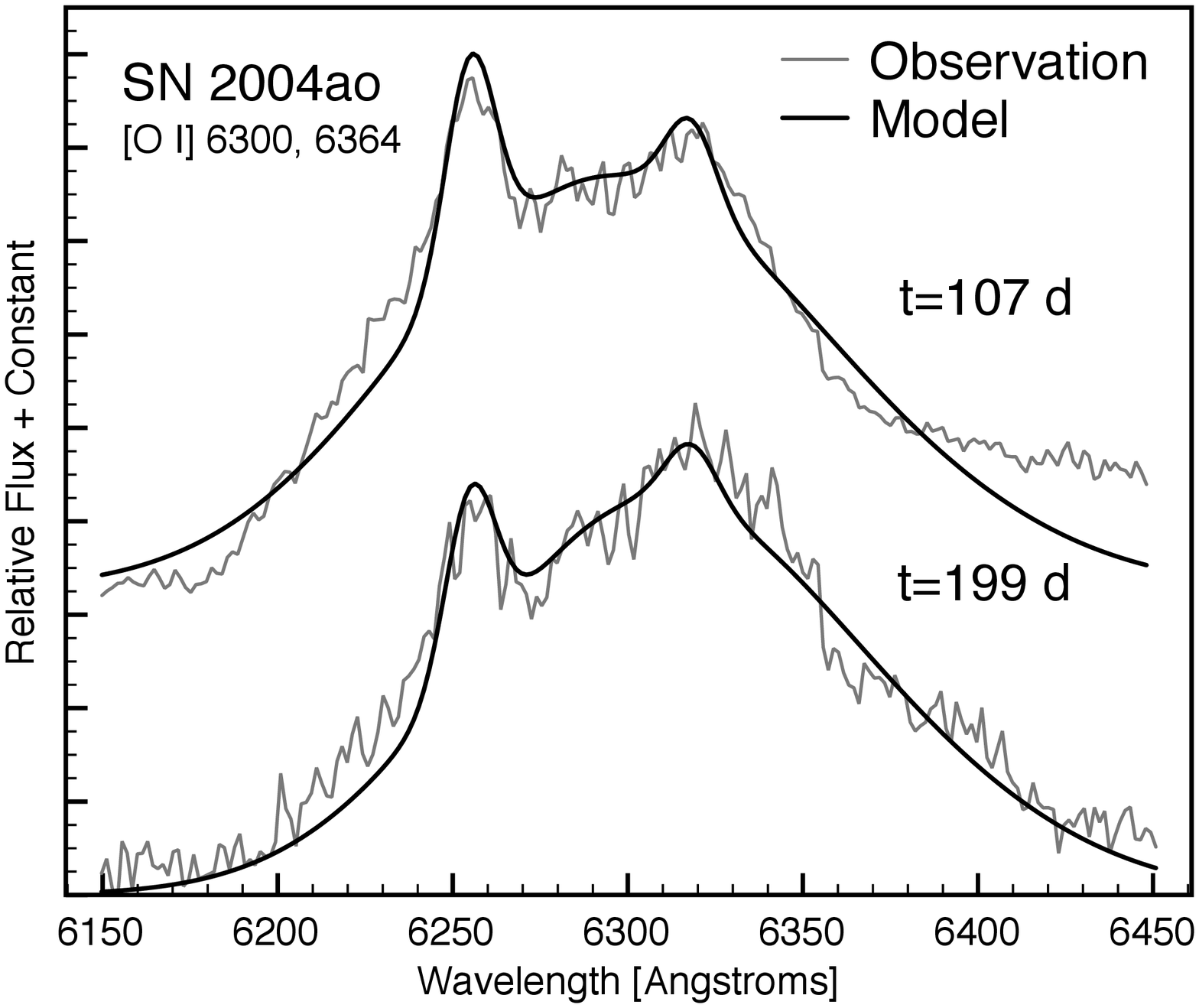} 
\caption{Line fitting models for the evolution of two symmetric profiles
following the decline of blueshifted emission peaks using the two-component
model illustrated in Figure 9.  Variables changed between epochs are relative
strength between narrow and broad components and the central wavelength of the
broad distribution. [\ion{O}{1}] 6300:6364 flux ratios of $\approx$2:1 and 3:1
were adopted for SN~2008ax and SN~2004ao, respectively. Slight changes to the
FWHMs of the distributions are introduced to compensate for changes in spectral
resolution.  Broad distributions originally blueshifted $\sim 20$ \AA\ at $t
\sim 100$ d gradually shift towards zero velocity by 200 d and the narrow
blueshifted components weaken in relative strength. Top panels show models for
all epochs, and bottom panels compare models and observed data for first and
last epochs. Observational data for SN~2004ao are from \citet{Modjaz08a}.} 
\end{figure*}

\subsubsection{A Single Blueshifted Emission Component?}

The [\ion{O}{1}] \dlambda 6300, 6364 emission-line profile that would follow
from a single preferentially blueshifted emission source was investigated
first. This configuration mimics emission originating from a blueshifted,
optically thick, and high density region (possibly the front side of a ring,
torus, or hollow shell with densities upward of $10^{10}$ cm$^{-3}$) where the
6300:6364 flux ratio could approach unity and the lack of a corresponding
redshifted emission peak might be due to significant internal extinction in the
ejecta. 

Both observed [\ion{O}{1}] $\lambda$5577 and \ion{Mg}{1}] $\lambda$4571
emission-line profiles were used as templates for the [\ion{O}{1}] \dlambda
6300, 6364 lines.  The [\ion{O}{1}] $\lambda$5577 profiles were used for
earlier epochs ($t \la 100$ d) and the \ion{Mg}{1}] $\lambda$4571 at later ones
($t \ga 200$ d), as these epochs represent when the lines are best observed.
The template profiles were separated by 3000 \kms\ in velocity, added, and the
result scaled to match the amplitude of the observed [\ion{O}{1}] \dlambda
6300, 6364 profile of the same epoch.  The SNe having the best time
coverage of our data set (SN~2008ax and SN~2004ao) were modeled at two epochs,
while two other objects displaying these lines with adequate S/N were modeled
at one epoch (SN 2004dk and SN 2006T). 

Figure 8 shows a comparison between the line fitting models and observed
[\ion{O}{1}] \dlambda 6300, 6364 profiles.  The top panels show our results
with the [\ion{O}{1}] $\lambda$5577 template. The best fits were obtained using
a 6300:6364 flux ratio of 1:1 in all three cases.  Noticeable disagreement on
the red side might be due to less reddening/extinction around 6300 \AA\ as
compared to around 5577 \AA, and/or contribution from other lines to the red of
the oxygen emission as evidenced by the moderate strength emission at higher
redward velocities.  The agreement is good for SN 2008ax and 2004ao, but rather
poor in SN 2006T where peak height is adequately matched but position and width
is not.

The bottom panels of Figure 8 present [\ion{O}{1}] line models using the
\ion{Mg}{1}] $\lambda$4571 line template. Both SN 2004ao and 2004dk show good
agreement and were best fit with templates added in a 3:1 ratio.  However, SN
2008ax was best fit with a ratio of 1.2:1.0 and showed an overall reasonable
fit but with noticeable disparity between the line model and observation around
zero velocity.

The results with our single component line fitting models corroborate
observations made in the \citet{Taubenberger09} survey of [\ion{O}{1}] \dlambda
6300, 6364 profiles of Ib/c SN.  Taubenberger et al.\ also used emission-line
profiles of \ion{Mg}{1}] $\lambda$4571 to model [\ion{O}{1}] \dlambda 6300,
6364 profiles as we have done here, and found general agreement between the two
profiles at sufficiently late phases. Furthermore, among the double-peaked
profiles they modeled, Taubenberger et al.\ reported that two SNe, SN
2000ew and 2004gt, could be fit very well with Gaussian distributions of the
deblended \dlambda 6300, 6364 lines originating from a single narrow,
blueshifted component.

Although the one-component line fitting models are encouraging in five out of
six profiles presented here, attributing double-peaked [\ion{O}{1}] emission to
the two doublet lines has problems. The most serious problem is the origin of
the preferentially blueshifted emission.  Expansion of SN ejecta should lower
density significantly such that emitted optical photons have minimal
interaction with the gas at nebular epochs \citep{Maeda08}, but the $\sim$1:1
flux ratio of the SN 2004ao and 2008ax [\ion{O}{1}] \dlambda6300, 6364 line
models and the lack of redshifted emission imply a significantly high optical
depth even at epochs of day $\sim$100 and greater. 
 
Another potential problem for the single component model is the evolution of
the blue/red peak height ratio. As seen in Figure 2, the blueshifted peak
decreases in strength relative to the redshifted peak with time.  The opposite
evolution is expected, however, if the two peaks are associated with the
[\ion{O}{1}] \dlambda 6300, 6364 doublet lines, since expansion of the
SN should drive the blue/red peak height ratio towards the nebular 3:1
ratio. A similar blue/red peak evolution away from the nebular 3:1 ratio in the
[\ion{O}{1}] \dlambda 6300, 6364 profile of SN 2004ao led \citet{Modjaz08a} to
conclude that the doublet lines were likely not behind the observed emission
peaks. 
 
However, the evolution of the [\ion{O}{1}] 5577 line profile in SN~2008ax
offers a possible explanation of the ratio change observed between emission
peaks.  Our multi-epoch observations show a gradual fading in the blue wing of
the [\ion{O}{1}] $\lambda$5577 profile (Figure 2, {\it left}), and our modeled
[\ion{O}{1}] profile shows that the bluest peak (representing the 6300 \AA\
component) sits on top of the blue wing of the 6364 \AA\ component.  Thus, if
the strength of the blue wing decays, the strength of the blue 6300~\AA\
component will drop relative to the red 6364~\AA\ one.  Unfortunately, attempts
at modeling this evolution at epochs beyond 74~d were inconclusive because of
the relatively poor S/N in the [\ion{O}{1}] $\lambda$5577 line at later epochs.

\subsubsection{Broad Emission Plus a Blueshifted Emission Component?}

We next investigated the [\ion{O}{1}] \dlambda 6300, 6364 line profile that
would follow from the contribution of two emission components.  Line fitting
models of [\ion{O}{1}] \dlambda 6300, 6364 emissions were constructed using the
IRAF package SPECFIT \citep{Kriss94}.  Each profile fit consisted of five
sources: (1) a linear continuum, (2) a {\it broad} (FWHM $\sim 6000$~\kms)
Gaussian centered at $6300 \pm 10$ \AA, (3) another Gaussian shifted 64 \AA\ to
the red fixed with the same FWHM but one third its strength, (4) a {\it narrow}
(FWHM $\sim 1000$~\kms) Gaussian at a blueshifted wavelength, and (5) another
narrow Gaussian 64 \AA\ to the red fixed with the same FWHM and one third its
strength.  The free parameters were the strength of the linear continuum, and
the position, FWHMs, and relative strengths of the pairs of peaks.  The line
profile model was tested with four SNe (SN~2007gr, SN~2008ax, SN~2008bo, and
SN~2004ao from \citealt{Modjaz08a}) showing symmetric and asymmetric profiles.

In Figure 9 we present the results of these reconstructed emission-line
profiles.  The top panels show fits for symmetric profiles.  The agreement
between our model profiles and the observational data for SN 2008ax and SN
2004ao is good.  SN 2006T, on the other hand, shows considerable disagreement.
The blue/red peak ratio is not matched, and there is considerable emission at
wavelengths longward of 6300 \AA.  We attempted to correct for the discrepancy
in the blue/red peak ratio using non-nebular 6300:6364 flux ratios (see the
inset of Figure 9). The model's agreement with the observed data for SN 2006T
improved as a result, although continued disagreement in the peak separation
and individual width was evident. 

The bottom panels of Figure 9 present our line fitting results for asymmetric
profiles.  In all cases the agreement is good,  although agreement is
relatively poor in SN~2007gr near the zero velocity peak. This discrepancy, as
well discrepancies in other line models, may in part be a reflection of the
inadequate use of Gaussian distributions to model the emission.  

The fixed 3:1 ratio used for the two components would seem to run counter to
the evolution of the blue/red peak height ratio observed in symmetric profiles.
However, \citet{Taubenberger09} recently concluded that the line centroids of
the bulk of [\ion{O}{1}] emission at phases earlier than day $\sim 200$  were
blueshifted $\sim$20 \AA, after which they drift close to zero velocity.  This
phenomenon offers an explanation for the apparent 6300:6364 ratio decrease over
observed epochs.  Assuming a broad/central + narrow/blueshifted interpretation,
a gradual shift toward zero velocity of the broad component would cause the
blueshifted $\lambda$6300 line of the narrow component to weaken in strength
relative to the $\lambda$6364 companion line.  We illustrate this possibility 
in Figure 10 for SN 2004ao and SN 2008ax.

While these profile fits are encouraging, two-component line fitting models
have notable shortcomings.  First, unlike the single-component scenario which
used genuine emission profiles of the SNe, these line models assume
Gaussian expansion profiles for both the broad and narrow components.  This
assumption along with the number of free variables used to replicate the
observed profiles and the possible redundancy in the reconstructions introduces
considerable uncertainty in the veracity of the model fits.  

Aside from these modeling concerns, there is also the question of why clumps
are mainly visible on the forward facing hemisphere, and rarely from the rear
side.  If clumps were randomly distributed, a variety of blueshifted and
redshifted emission peaks would be expected. Any explanation in which a
preferential visibility of blueshifted peaks might be caused by high internal
extinction is complicated by the fact that considerable broad emission is seen
at both higher blueshifted and redshifted velocities. 

\section{Discussion and Conclusions}

Double-peaked [\ion{O}{1}] \dlambda 6300, 6364 emission-line profiles are not
uncommon in the late-time spectra of stripped-envelope CCSNe and have been
interpreted as possibly originating from ejecta arranged in a torus or
disk-like geometry.  As part of an investigation into this phenomenon, we
obtained optical spectra of five SNe (SN~2007gr, SN~2007rz, SN~2007uy,
SN~2008ax, and SN~2008bo) and examined the emission-line profiles of
[\ion{O}{1}] $\lambda$5577, [\ion{O}{1}] \dlambda 6300, 6364, \ion{O}{1}
$\lambda$7774, and \ion{Mg}{1}] $\lambda$4571.  Because our spectra of
SN~2008ax covered the broadest time interval (day 67 to 418) and showed strong,
sharp, and well defined [\ion{O}{1}] emission peaks, we used SN~2008ax to help
guide the investigation.  We also examined the late-time [\ion{O}{1}] \dlambda
6300, 6364 line profiles of 13 other stripped-envelope CCSNe taken from the
literature.

\subsection{Empirical Results of Our Study}

Our investigation into the properties of late-time [\ion{O}{1}] \dlambda 6300,
6364 line profiles of stripped CCSNe exhibiting double-peaked line profiles
showed the following:

-- Doubled-peaked line profiles are primarily in [\ion{O}{1}] \dlambda 6300,
6364 emission (Figure 4 and 5) and can be categorized into two types.  One type
shows conspicuous {\it symmetric} emission peaks positioned about 6300 \AA\
that are persistent across the entire observing period.  The other type shows
{\it asymmetric} emission profiles with one emission peak near zero velocity
(i.e., 6300 \AA) plus one or more peaks at blueshifted velocities that
sometimes persist hundreds of days post outburst or evolve over short ($t < 2$
months) timescales.

-- Emission peaks in symmetric double-peaked [\ion{O}{1}] \dlambda 6300, 6364
line profiles are often close to $3000$ \kms\ apart (i.e., 64 \AA; Figure 7),
nearly the 63.5 \AA \  wavelength spacing of the [\ion{O}{1}] doublet,
suggesting the double-peaks are simply reflecting the doublet nature of the
[\ion{O}{1}] line emission. Over time, the velocities of the two peaks do not
change, but the blue/red peak intensity ratio slowly decreases.  

-- Conspicuous redshifted emission peaks are not observed in asymmetric profile
cases. As seen for symmetric profiles, the position of blueshifted peaks does
not change between observed epochs, but their strength relative to emission
near zero velocity tends to weaken with time.

-- When double-peaked [\ion{O}{1}] profiles are present, single-peaked
emission peaks at blueshifted velocities matching blueshifted peaks in
[\ion{O}{1}] \dlambda 6300, 6364 are often seen for [\ion{O}{1}] $\lambda$5577.
The \ion{Mg}{1}] $\lambda$4571 and \ion{O}{1} $\lambda$7774 lines generally
show more blueshifted emission than redshifted emission and sometimes evidence
of peaks at velocities matching blueshifted and/or zero-velocity peaks in the
[\ion{O}{1}] \dlambda 6300, 6364 lines.

\subsection{Results of Double-peaked [\ion{O}{1}] Line Profiles Fits}

The high incidence of $\approx$64 \AA\ separation between emission peaks of
symmetric profiles plus the lack of redshifted emission peaks in asymmetric
profiles suggests that emission from the rear of the SN may be suppressed.
This leads us to conclude that double-peaked [\ion{O}{1}] \dlambda 6300, 6364
line profiles of some stripped-envelope, core-collapse SNe are not necessarily
signatures of emission from the front and rear faces of a torus or elongated
shell of O-rich ejecta as has been proposed.

Alternative interpretations for the observed [\ion{O}{1}] \dlambda 6300, 6364
profiles were investigated through line-fitting models.  Two models were
explored: Model 1 where the [\ion{O}{1}] profile arises only from
preferentially blueshifted emission, and Model 2 where the [\ion{O}{1}] profile
consists of two separate emission components, a broad emission source centered
around zero velocity and a narrow, blueshifted source.  Both line-fitting
models reproduced observed [\ion{O}{1}] \dlambda 6300, 6364 profiles in the
majority of test cases. Model 2 showed better overall agreement to the data for
all six cases of symmetric and asymmetric profiles across all epochs
investigated, but Model 1 had fewer parameters and convincing results
in a subset of the symmetric profiles.  Similar line-fitting results have been
reported in \citet{Taubenberger09}. 

It should be noted that a torus or elongated shell model is still viable
despite the concerns raised above if the rear portion of an O-rich torus or
shell is somehow hidden by scattering or dust.  Moreover, two of the 18 SNe
studied in this sample, SN 2003jd and 2006T, exhibited both blueshifted and
redshifted emission peaks with separations and widths that could not be modeled
under the alternative interpretations offered here. Hence, these objects
represent notable exceptions to the concerns raised in this paper and remain
strong candidates for a torus geometry of the ejecta as previously proposed. 

\subsection{Outstanding Questions}

Two questions follow from our observations and findings: 

{\it Why do these emission-line profiles lack redshifted emission
peaks?}   The predominance of blueshifted emission features in the profiles
studied is unclear but internal scattering or dust obscuration of emission
from far side ejecta are the most likely scenarios.  \citet{Taubenberger09},
who also found predominantly blueshifted peaks in [\ion{O}{1}] \dlambda 6300,
6364 emission-line profiles in a sample of 39 Type Ib/c SNe, favored an opaque
inner ejecta scenario, citing that other explanations such as ejecta geometry,
dust formation, and contamination from other lines, could not account for all
observed trends.   

{\it What is the physical nature of these blueshifted features?} The variety
and strength of the emission peaks are suggestive of asphericity in the ejecta.
But whether these features can be conclusively associated with cones, jets, or
tori of unipolar/bipolar explosions is uncertain and beyond the scope of this
paper.  We note that the difference between symmetric and asymmetric profiles
in our two-component line-fitting models is a consequence of the parameters
adopted for the narrow, blueshifted component (i.e., FWHM, central velocity,
strength relative to the broad component near zero velocity, and 6300:6364 flux
ratio).  Sophisticated models incorporating radiative transfer effects could
explore how these line-fitting parameters may be related to physical quantities
such as mass, velocity, size, and density of the O-rich material, as well as
changes with viewing angle. Such models could also investigate specific
extinction mechanisms of the suspected opaque inner region.  

\subsection{Future Observations} 

Much more robust tests of line-fitting models are possible with spectra of
improved time coverage, spectral resolution, and S/N for those CCSNe displaying
double-peaked [\ion{O}{1}] \dlambda 6300, 6364 profiles.  Moreover,
observations optimized around the lines of \ion{Mg}{1}] $\lambda$4571,
[\ion{O}{1}] $\lambda$5577, and \ion{O}{1} $\lambda$7774 at late epochs
starting from 50 days past maximum light could greatly help test the trends
observed in our sample and help eliminate one of the two line models (single or
two component) discussed here. Infrared observations might also place
constraints on the cause of the suspected internal extinction.  While an opaque
inner region appears to best explain the lack of redshifted features, late-time
infrared studies of other stripped-envelope CCSNe could characterize the
presence of internal molecules and dust as a test of one source of extinction
at the epochs studied here.

\acknowledgements

This work has made use of data from many previous observers which proved
invaluable in our analysis. We thank M.\ Modjaz for many helpful comments and
kindly providing spectra for SN~2004ao, SN~2006T, SN~2005bf, SN~2004gt,
SN~2004dk, and SN~2008D.  We also thank P.\  H\"{o}flich for additional helpful
discussions, and the referee for constructive suggestions.  Some data presented
here were obtained at the MMT Observatory, a joint facility of the Smithsonian
Institution and the University of Arizona.  Supernova research at the Harvard
College Observatory is supported by the National Science Foundation through
grants AST-0606772 and AST-0907903. D.M.'s research was supported in part by a
Canadian NSERC award.

\end{document}